\documentclass{pspum-l}
\usepackage{color}
\usepackage{amscd}
\usepackage[psamsfonts]{amssymb}
\usepackage{transfig,amssymb,amsmath}

\newtheorem{thm}{Theorem}[section]
\newtheorem{lem}[thm]{Lemma}
\newtheorem{propos}[thm]{Proposition}

\theoremstyle{definition}
\newtheorem{defin}[thm]{Definition}

\theoremstyle{remark}
\newtheorem{remark}[thm]{Remark}

\numberwithin{equation}{section}

\def\ali{\hfill\break}
\def\w{\omega}

\def\unN{\{1\ldots ,N\}}
\def\rrr{{\mathcal R}}
\def\demi{{1\over 2}}
\def\ddd{{\mathcal D}}
\def\ppp{{\mathcal P}}

\def\lll{{\BL}}

\def\lllG{{\BL_F^G}}

\def\aaa{{\mathcal A}}

\def\eee{{{\mathcal E}}}
\def\nn{{<n>}}
\def\ND{{ND}}
\def\infi{{<\infty>}}

\def\unN{{\{1,\ldots ,N\}}}

\def\j1p0{{j_1^0,\ldots ,j_p^0}}

\def\jp1{{j_p,\ldots ,j_1}}
\def\j1p{{j_1,\ldots ,j_p}}

\def\BR{{\Bbb R}}
\def\supp{{\hbox{supp}}}
\def\BC{{\Bbb C}}
\def\BN{{\Bbb N}}
\def\BZ{{\Bbb Z}}

\def\BL{{\Bbb L}}

\def\BP{{\Bbb P}}

\def\nonu{\nonumber}

\def\oeta{{\overline \eta}}

\def\ord{{\hbox{ord}}}
\def\beq{\begin{eqnarray}}
\def\eeq{\end{eqnarray}}
\def\beqn{\begin{eqnarray*}}
\def\eeqn{\end{eqnarray*}}
\def\bpr{\begin{propos}}
\def\epr{\end{propos}}

\def\loc{{\hbox{loc}}}

\def\bF{{\partial F}}
\def\rF{{F/\rrr}}
\def\bQ{{Q_{\partial F}}}
\def\symF{{\hbox{Sym}_F}}

\def\sym{{\hbox{Sym}}}

\def\11{{(1,1)}}

\def\Id{{\hbox{Id}}}

\def\im{\hbox{Im}}

\def\ksi{{\xi}}
\def\oksi{{\overline \xi}}
\def\vect{{\hbox{vect}}}

\def\Qu01{{Q^{u_0,u_1}}}

\def\perpw{{\perp_{\omega}}}
\def\perps{{\perp_{<,>}}}
\def\vect{{\hbox{Span}}}
\def\un{{<1>}}
\def\pF{{\partial F}}
\def\unK{{\{1,\ldots ,K\}}}
\def\build#1#2#3{\mathrel{\mathop{\kern 0pt #2}\limits_{#1}^{#3}}}

\begin{document}
\title[Renormalization Map of Self-Similar Lattices]{Electrical Networks,
Symplectic Reductions, and Application to the Renormalization Map
of Self-Similar Lattices}
\author{Christophe Sabot}
\address{CNRS UMR 7599, Laboratoire de probabilit\'es et mod\`eles al\'eatoires,
Universit\'e Paris 6, 4, Place Jussieu, 75252 Paris cedex 5,
France, and DMA, Ecole Norm. Sup., 45 rue d'Ulm, 75005 Paris,
France.}
%\curraddr{
%Ecole Normale Sup\'erieure,
%DMA, 45, rue d'Ulm, 75005 Paris, France
%}
\email{sabot@ccr.jussieu.fr}

\thanks{The author was supported by CNRS}

%    General info
\date{Mars 2003 and, in revised form, October 2003.}


\begin{abstract}
The first part of this paper deals with electrical networks and
symplectic reductions. We consider two operations on electrical
networks (the ``trace map" and the ``gluing map") and show that they
correspond to symplectic reductions. We also give several general
properties about symplectic reductions, in particular we study the
singularities of symplectic reductions when considered as rational
maps on Lagrangian Grassmannians. This is motivated by
\cite{Sabot5} where a renormalization map was introduced in order
to describe the spectral properties of self-similar lattices. In
this text, we show that this renormalization map can be expressed
in terms of symplectic reductions and that some of its key
properties are direct consequences of general properties of
symplectic reductions (and the singularities of the symplectic
reduction play an important role in relation with the spectral
properties of our operator). We also present new examples where we
can compute the renormalization map.
\end{abstract}

\maketitle

\section*{Introduction}

In \cite{Sabot5}, we introduced a renormalization map in order to
describe the spectral properties of Laplace operators on finitely
ramified self-similar lattices. This map is rational and defined
on a Lagrangian Grassmannian. The aim of this text is to present
this map from a different point of view. We insist on the aspects
of symplectic geometry, and in particular on the role played by
symplectic reductions. In this respect, we take inspiration from
the works of Colin de Verdi\`ere (\cite{Colin1}, \cite{Colin2}),
where Lagrangian compactifications and symplectic reductions are
related to operations on electrical networks. One of the main
goals of this text is to show that the crucial properties of the
map introduced in \cite{Sabot5} are consequences of general
properties of symplectic reductions. These properties, which seem
to be new, are proved in section 3 and essentially concern the
singularities of the symplectic reduction, when considered as a
rational map. We also show that the symplectic reduction can be
lifted, in a natural way, to a linear map, through the Pl\"ucker
embedding. This is a key feature in \cite{Sabot5}.
 This leads us to introduce a
class of rational maps, which is a natural generalization of the
maps that appear in the context of self-similar structures, cf.
section 6.

We also present new examples, in particular an example related to
the spectrum of Schreier graphs of some automatic groups (cf.
\cite{Grig1}, \cite{Grig2}), and explain, with more detail than in
\cite{Sabot5}, how to explicitly compute the renormalization map
in the ``nested fractal like" cases (which have many symmetries).
 \ali

Let us now briefly explain the main ideas of this work. In
sections 1 and 2, we are concerned about two operations on
electrical networks and their translation in terms of symplectic
reductions. If $F=\{1,\ldots , K\}$ is a finite set, an electrical
network on $F$, is a family of non-negative reals
$(\rho_{i,j})_{i,j\in F,\; i\neq j}$, such that
$\rho_{i,j}=\rho_{j,i}$, and a family of non-negative reals
$(\rho_i)_{i\in F}$. The $\rho_{i,j}$ are called the conductances,
and the $\rho_i$ are the dissipative terms. We say that the
network is conservative when $\rho_i=0$ for all $i$ in $F$. For
any function $f:F\rightarrow \BR$ (which represents the potential
on the poles of $F$), the energy dissipated by the network is
$$
\eee^\rho (f,f)=\demi  \sum_{i,j\in F}\rho_{i,j}
(f(i)-f(j))^2+\sum_{i\in F}\rho_i (f(i))^2.
$$
(We denote by $\eee^\rho(f,g)$ the bilinear form obtained by
polarization). For any potential $f:F\rightarrow \BR$, the current
is the element $I^\rho_f$ of the dual space $(\BR^F)^*$, defined by
$$
I^\rho_f(h)=\eee^\rho (f,h), \;\;\; \forall h\in \BR^F.
$$
Physically, $I^\rho_f$ represents the current flowing through the
poles of $F$, when the potential $f$ is imposed on $F$.

In section 2, we consider two natural operations on electrical
networks. The first one is the so-called ``trace map" (this
terminology comes from Dirichlet forms). One way to present this
operation is the following. Consider a subset $\partial F$ of $F$
($\partial F$ is often viewed as a boundary set for $F$, which
justifies this notation), then there is a unique electrical
network on $\pF$, denoted $(\rho_{i,j}^{\partial F})_{i,j \in
\pF,\; i\neq j}$, $(\rho^{\pF}_i)_{i\in \pF}$, such that for any
function $f:\pF\rightarrow \BR$,
$$
\eee^{\rho^{\pF}}(f,f)=\inf_{{h:F\rightarrow \BR\atop h_{|\pF}=f}}
\eee^{\rho}(h,h).
$$
When the electrical network $\rho$ is irreducible (cf. section 1),
this infimum is attained at a unique point, denoted $Hf$, which is
the harmonic extension of $f$ with respect to $\eee^\rho$.
Physically, $\eee^{\rho^{\pF}}(f,f)$ is the energy dissipated by
the network when the potential $f$ is imposed on the poles of
$\pF$. It is clear that the current $I^\rho_{Hf}$ induced by the
potential $Hf$ is supported by $\pF$.

The second operation we consider is the following: suppose that
$(F,\rho)$ is an electrical network and that $\rrr$ is an
equivalence relation on $F$. There is a natural way to define an
electrical network $\rho^\rrr$ on the quotient set $F/\rrr$ by
$$
\rho^{F/\rrr}_{x,y}=\sum_{{i,j\in F\atop \pi( i)=x,\; \pi(j)=y}}
\rho_{i,j}, \;\;\; \rho^{F/\rrr}_x=\sum_{i\in F, \; \pi(i)=x} \rho_i,
$$
where $\pi$ is the canonical surjection $\pi: F\rightarrow
F/\rrr$.

For some reason that will appear later, it is important to define
these two maps not only on electrical networks, but on the larger
set of symmetric matrices. Let us denote by $\symF$ the space of
$F\times F$ symmetric matrices (we take the coefficients
in $\BR$ in this introduction). For an electrical network $\rho$,
we denote by $Q_\rho$ the element of $\symF$ defined by
$$
\eee^\rho(f,h)=\; <Q_\rho f,h>, \;\;\; \forall f,h\in \BR^F,
$$
(where $<\cdot ,\cdot>$ is the canonical scalar product on
$\BR^F$). It is clear that $Q_\rho$ determines completely $\rho$,
and that the subset $\{Q_\rho, \;\rho\hbox{ elec. net.}\}$ is a
cone of $\symF$, with non-empty interior. The maps $\rho\mapsto
\rho^{\pF}$ and $\rho\mapsto \rho^{F/\rrr}$, naturally induce two
maps from the cone $\{Q_\rho\}$ to respectively $\sym_\pF$ and
$\sym_{F/\rrr}$. Furthermore, it is easy to check that the
coefficients of $Q_{\rho^\pF}$ and $Q_{\rho^{F/\rrr}}$ are
rational in the coefficients of $Q_\rho$ (cf. section 2.1). Hence,
we can extend these maps into rational maps on $\sym_F$, that we
denote by
$$
\begin{array}{rcl}
\symF & \rightarrow &\sym_\pF
\\
Q&\mapsto &Q_\pF
\end{array} \;\;\hbox{ and }\;\;
\begin{array}{rcl}
\symF & \rightarrow &\sym_{F/\rrr}
\\
Q&\mapsto &Q_{F/\rrr}
\end{array}.
$$
(Explicit expressions for these maps are given in section 2).

An electrical network can be considered as a Lagrangian subspace.
Let us consider $V_F=\BR^F\oplus (\BR^F)^*$, where $(\BR^F)^*$ is
the dual space of $\BR^F$. We consider the bilinear symplectic
form on $V_F\times V_F$, defined by
$$
\w ((x,\ksi),(x',\ksi'))=\ksi'(x)-\ksi(x'),
$$
for any $(x,\ksi)$ and $(x',\ksi')$ in $V_F\simeq \BR^F\times
(\BR^F)^*$. Let $W$ be a linear subspace of $V_F$. We denote by
$W^o$ the orthogonal of $W$ with respect to $\w$. By definition,
the subspace $W$ is isotropic if $W\subset W^o$, and coisotropic
if $W^o\subset W$. A Lagrangian subspace is a maximal isotropic
subspace of $V_F$ (which, thus is also coisotropic and of
dimension $K=\vert F\vert$). We denote by $\lll_F$ the set of
Lagrangian subspaces of $V_F$. The set $\lll_F$ has the structure
of a smooth projective variety of dimension $\dim \symF=K(K-1)/2$
(cf. section 1.2). If $W$ is a coisotropic subspace, then the
symplectic form $\w$ induces a symplectic form on $W/W^o$, and if
$L$ is a Lagrangian subspace of $V_F$, then $(L\cap W)/W^o$ is a
Lagrangian subspace of $W/W^o$ (cf. section 1.3). The symplectic
reduction is defined as the map $t_W: \lll_F\rightarrow
\lll_{W/W^o}$ given by $t_W(L)=(L\cap W)/W^o$ (where
$\lll_{W/W^o}$ is the variety of Lagrangian subspaces of $W/W^o$).
The map $t_W$ is defined everywhere on $\lll_F$, but is not
everywhere smooth. The singularities of this map play an important
role in relation with the operations on electrical networks we
have described (and section 3 of this paper is devoted to the
study of the singularities of the symplectic reduction, when
considered as a rational map).

With any electrical network $\rho$, we associate the subspace
$$
L_\rho =\{f+ I_f^\rho, \; f\in \BR^F\} \subset V_F,
$$
which is a Lagrangian subspace of $V_F$ and determines $\rho$
completely. More generally, if $Q$ is in $\symF$, then we can
define the subspace
$$
L_Q=\vect \{ e_i+ (\sum_{j=1}^K Q_{i,j} e_j^*)\}_{i=1,\cdots
,K},
$$
which is a Lagrangian subspace of $V_F$ ($(e_i)$ is the canonical
basis of $\BR^F=\BR^K$, and $(e_i^*)$ the dual basis). It is clear
with these notations that $L_\rho=L_{Q_\rho}$. The map $Q\mapsto
L_Q$ defines an embedding of $\symF$ into the variety $\lll_F$,
such that $\lll_F\setminus \symF$ is the a subvariety of $\lll_F$
of codimension 1 given by
$$
\lll_F\setminus \symF =\{L\in \lll_F, \;\; L\cap (0\oplus
(\BR^F)^*)\neq \{0\}\}.
$$
Hence, $\lll_F$ defines a compactification of $\symF$ (which is in
general different
from the compactification by the projective space of
dimension $K(K-1)/2$).

Let us come back to the operations of restriction and gluing we
have defined. The trace map $Q\mapsto Q_{\pF}$ and the gluing map
$Q\mapsto Q_{F/\rrr}$, naturally induce the maps $L_Q\mapsto
L_{Q_\pF}$ and $L_Q\mapsto L_{Q_{F/\rrr}}$ on $\symF\subset
\lll_F$. The main point of section 2, is to show that these two
maps coincide with symplectic reductions. More precisely, this
means that we can find  some explicit coisotropic subspaces of
$V_F$, $W_\pF$ and $W_{F/\rrr}$, such that $W_\pF/ (W_\pF)^o\simeq
V_{\pF}$ and $W_{F/\rrr}/(W_{F/\rrr})^o\simeq V_{F/\rrr}$ and such
that $t_{W_\pF}(L_Q)= L_{Q_{\pF}}$, and $t_{W_{F/\rrr}}(L_Q)=
L_{Q_{F/\rrr}}$ (for the trace map, this was proved by Colin de
Verdi\`ere in \cite{Colin1}). The main interest of these formulas
is to give an explicit expression of the extension of the trace
map and the gluing map to the Lagrangian compactification
$\lll_F$.

This work is motivated by the spectral analysis of self-similar
lattices. Let us describe in this introduction the simple case of
the Sierpinski gasket. Let $F=F_{<0>}$, be the set of vertices of
a regular triangle, and $F_{<1>}$ be the (non-disjoint) union of 3
copies of $F$, as shown on figure 1. Formally, it means that
$F_{<1>}=\{1,2,3\}\times F/\rrr$, where $\rrr$ is a certain
equivalence relation which represents the connexions in $F_{<1>}$.
We denote by $\pF_{<1>}\simeq F$, the boundary points of $F_{<1>}$
(the circled points on figure 1). Then, $F_{<2>}$ is constructed
as 3 copies of $F_{<1>}$, glued together by the boundary points
$\pF_{<1>}$, as shown on figure 1.
 The boundary
set of $F_{<2>}$, $\pF_{<2>}$, is the set of circled points on
figure 1. \ali \ali \centerline{\input{figure1-new.pstex_t}}
 \ali

Repeating this operation, we construct a sequence of lattices
$F_\nn$, together with their boundary sets $\partial F_\nn$
(consisting of the 3 vertices of the larger triangle). Let us now
consider an electrical network $\rho$ on $F$. Then, we can
naturally define an electrical network $\rho_\nn$ on $F_\nn$:
$\rho_{<1>}$ is constructed from $\rho$ by first making three
copies of $(F,\rho)$, and then $(F_{<1>},\rho_{<1>})$ is obtained
by the gluing procedure described at the beginning of the
introduction (considering that $F_\un$ is a quotient of
$\{1,2,3\}\times F$). The electrical network $\rho_{<n+1>}$ is
defined similarly from $\rho_\nn$. If $b$ is a positive measure on
$F$, then we can construct a self-similar positive measure $b_\nn$
on $F_\nn$ in a natural way (the details are in section 4).
 Let $H_\nn$ be the self-adjoint operator on
$L^2(b_\nn)$ defined by
$$
\eee^{\rho_\nn} (f,h)= -\int_{F_\nn} (H_\nn f) h db_\nn(x), \;\;\;
\forall f,h \in \BR^{F_\nn}.
$$
The operator $H_\nn$ is a self-similar Schr\"odinger operator on
the sequence of self-similar lattices $F_\nn$ ($H_\nn$ is of
``Laplace type" when $\rho$ is conservative). In \cite{Sabot5},
\cite{Sabot6}, and in this work we are interested in the spectral
properties of this operator. Remark that $f:{F_\nn}\rightarrow
\BR$ is an eigenvalue of $H_\nn$, with eigenvalue $\lambda$,
 if and only if
$$
\eee^{\rho_\nn} (f,h)=-\lambda \int_{F_\nn} fh db_\nn, \;\;\;
\forall h \in \BR^{F_\nn}.
$$
As shown in \cite{Sabot5},
the spectral properties of $H_\nn$ are related to the dynamics of
a certain renormalization map that we describe now. Let us denote
by $\tilde F_{<1>}=\{1,2,3\}\times F$, three copies of $F$. If
$Q$ is a symmetric $F\times F$ matrix, then $\tilde Q_{<1>}$ is
defined as the block diagonal
 $\tilde F_{<1>}\times \tilde F_{<1>}$ matrix
obtained by making three copies of $Q$ on each subset $\{i\}\times
F\subset \tilde F_\un$. Then $Q_{<1>}$ is the element of
$\sym_{F_{<1>}}$ obtained from $\tilde Q_{<1>}$ by the gluing map
we have described. Then we define $TQ$ as the element of
$\sym_{\pF_{<1>}}$ obtained by the trace map:
$$
TQ=(Q_{<1>})_{\pF_{<1>}}.
$$
Since there is a natural identification between $\pF_{<1>}$ and
$F$, we see that $T$ is a map from $\symF$ to $\symF$ (the
coefficients of $TQ$ are rational in the coefficients of $Q$). We
see that the map $T$ is the composition of three maps
$$
T:Q\build{\scriptstyle copies}{\longmapsto}{} \tilde Q_{<1>}
\build{\scriptstyle gluing}{\longmapsto}{} Q_{<1>}
\build{\scriptstyle trace map}{\longmapsto}{}
TQ=(Q_{<1>})_{\pF_{<1>}}.
$$
The last two operations correspond to symplectic reductions on the
Lagrangian compactification. Since a composition of two symplectic
reductions is a symplectic reduction (cf. section 1.3), we see
that the extension of the map $T$ to the Lagrangian
compactification $\lll_F$ has the following simple expression
\begin{equation}\label{intro.0}
\begin{array}{lccccl}
g:
&\lll_F & \build{\scriptstyle copies}{\longrightarrow}{}
 & \lll_{\tilde F_{<1>}} &
\build{\scriptstyle {{symplectic\atop  reduction}}}{\longrightarrow}{}
&\lll_F,
\\
&L& {\longmapsto } &\tilde L_{<1>}& {\longmapsto }& g(L)=
t_{W_\un}(\tilde L_{<1>})
\end{array}
\end{equation}
where $W_{\un}$ is a certain coisotropic subspace of $V_{\tilde
F_{<1>}}$ (which is made explicit in section 4). This
renormalization map is crucial in the understanding of the
spectral properties of the operator $H_\nn$; in particular it is
crucial to understand the behavior of
$$
g^n(L_{Q_\rho +\lambda I_b})
$$
where $I_b$ is the diagonal  $F\times F$ matrix  with diagonal
terms $(I_b)_{x,x}=b(x)$. The reason is that $g^n(L_{Q_\rho
+\lambda I_b})$ is equal to the following Lagrangian subspace of
$V_{\pF_\nn}\simeq V_F$: consider the functions
$f:\BR^{F_\nn}\rightarrow \BR$ such that
\begin{eqnarray}
\label{intro.1} \;\;\;\;\;\;\;\;\;\; \eee^{\rho_\nn}(f,h)+\lambda
\int_{F_\nn} fh db_\nn =0, \;\;\; \forall h\in \BR^{F_\nn} \;
\hbox{ s.t. }\; h_{|\pF_\nn}=0.
\end{eqnarray}
For such a function we denote by $I^{\rho_\nn, \lambda}_f$ the
element of $(\BR^{F_\nn})^*$ such that
$$
I_f^{\rho_\nn,\lambda}( h)= \eee^{\rho_\nn}(f,h)+\lambda \int_{F_\nn}
fh db_\nn , \;\;\; \forall h\in \BR^{F_\nn}.
$$
By (\ref{intro.1}), $I_f^{\rho_\nn,\lambda}$ is supported by
$\pF_\nn$ and hence lies in $(\BR^{\pF_\nn})^*$. Then,
\begin{eqnarray}\label{intro.2}
g^n(L_{Q_\rho +\lambda I_b}) =\left\{ f_{|\pF_\nn}+I_f^{\rho_\nn,
\lambda}, \;\; f \; \hbox{ solution of (\ref{intro.1})}\right\}.
\end{eqnarray}
\begin{remark}
Otherwise stated, it means that we consider the solutions of
$(H_\nn-\lambda)f=0$ on $F_\nn\setminus \pF_\nn$, and that $I_f^{\rho_\nn,\lambda}$
plays the role of a kind of discrete derivative on
$\pF_\nn$. Hence, $g^n(L_{Q_\rho+\lambda I_b})$
is the subspace generated by the boundary values of the space of solutions
of $(H_\nn-\lambda)f=0$ on $F_\nn\setminus \pF_\nn$.
\end{remark}
 This formula is useful to understand the role played by the
renormalization map $g$. Indeed, if $f$ is
an eigenfunction of $H_\nn$ with eigenvalue $\lambda$,
we see that it is a solution
of (\ref{intro.1}) with $I_f^{\rho_\nn,\lambda}=0$.
Hence, if $f_{|\pF_\nn}\neq 0$, it means that
\begin{eqnarray}
\label{intro.3}
g^n(L_{Q_\rho +\lambda I_b})\cap(\BR^F\oplus 0)
\end{eqnarray}
is a non trivial subspace of $V_F$. Similarly, the intersection
\begin{eqnarray}
\label{intro.4}
g^n(L_{Q_\rho +\lambda \Id})\cap(0\oplus (\BR^F)^*)
\end{eqnarray}
is related to the Dirichlet eigenfunctions of $H_\nn$, with
eigenvalues $\lambda$. Hence, if $C^+=\BR^F\oplus 0$ and
$C^-=0\oplus (\BR^F)^*$, we see that the Neumann (resp. Dirichlet)
spectrum is related to the intersection of the curve
$\lambda\mapsto L_{Q_\rho +\lambda I_b}$ with the hypersurface
$f^{-n}(C^+)$ (resp. $f^{-n}(C^-)$). Technically, to count these
eigenvalues with multiplicities, we consider the current of
integration on $C^+$ (resp. $C^-$) and its pull-back by $f^n$ (cf.
section 4.6).

The last point we want to insist on in this introduction deals
with the relation between the singularities of the renormalization
map $g$ and a certain type of eigenfunctions on $F_\nn$. These
special eigenfunctions are the so-called ``Neumann-Dirichlet"
eigenfunctions (N-D for short): a function $f:F_\nn\rightarrow
\BR$ is a N-D eigenfunction with eigenvalue $\lambda$, if
$$
H_\nn f=\lambda f, \;\;\;\hbox{and}\;\;\; f_{|\pF_\nn}=0.
$$
Hence, $f$ is an eigenfunction with both Neumann and Dirichlet
boundary conditions (and actually, with any mixed boundary
condition). Remark now that the boundary  values (i.e.
$f_{|\pF_\nn}$ and $I_f^{\rho_\nn,\lambda}$) of these
eigenfunctions vanish, and thus do not contribute to the
Lagrangian subspace $g^n(L_{Q_\rho +\lambda I_b})$ (cf. formula
(\ref{intro.2})). Actually, these eigenfunctions appear, with
multiplicities, as the singularities of the map $g$ and its
iterates $g^n$. This was proved in \cite{Sabot5}, but in this
text, we clarify this point by a systematic analysis of the
singularities of symplectic reductions.

These are the main ideas underlying this work. Part of them were
already presented in \cite{Sabot5}, but compared to \cite{Sabot5},
the main goals are
\begin{itemize}
\item
To explain the relations between operations on electrical networks
and symplectic reductions (sections 1 and 2).
\item
To describe the singularities of symplectic reductions. We also
give several general results about symplectic reductions, which
are the bases of some of the key properties of our renormalization
map. In particular, we show that symplectic reductions can be
lifted to the exterior product $\bigwedge^K V$ by a linear map,
using the Pl\"ucker embedding of $\BL_V$ into the projective space
$\ppp(\bigwedge^K V)$. This generalizes to symplectic reductions
one of the main arguments of \cite{Sabot5}. This is done in
section 3. Let us stress that this section is more or less
self-contained and does not appeal to such notions as electrical
networks or self-similar lattices.
\item
We present the renormalization map introduced in \cite{Sabot5}
from the point of view of symplectic geometry. More precisely, we
give an explicit expression of the renormalization map on the
Lagrangian compactification in terms of symplectic reduction. This
is new compared to \cite{Sabot5}. We also use several general
results obtained in section 3, to recover some of the key results
of \cite{Sabot5}. This is done in section 4.
\item
In section 6, we propose  a class of rational maps on Lagrangian
Grassmannians with a simple and natural definition (which
essentially reproduces the figure (\ref{intro.0})) and which
shares the same basic properties as the renormalization maps of
self-similar lattices.
\item
Finally, we present some new examples (cf. \cite{Sabot-review1}
for other examples). In particular, we show that
one of the rational maps appearing in relation with some automatic
groups in the works of Grigorchuk, Bartholdi and Zuk (cf.
\cite{Grig1}, \cite{Grig2}) can be handled in our framework
(section 7). In section 7, we also try to clarify how to proceed
to make explicit computations when the structure has a large group
of symmetries.
\end{itemize}

\section{Electrical networks, Lagrangian compactification and
Pl\"ucker embedding}
\subsection{Electrical networks}
Let $F=\unK$ be a finite set. We denote by $\symF(\BC)$,
$\symF(\BR)$ (or $\sym_K(\BC)$, $\sym_K(\BR)$) the set of
symmetric $K\times K$ matrices with coefficients in $\BR$ or
$\BC$. By abuse of notation, we identify a $K\times K$ matrix with
the linear operator induced on $\BR^F$ or $\BC^F$.

We call dissipative electrical network a family $(\rho_{i,j})$,
$i\neq j$, $i,j\in F$, and a family $(\rho_i)$,
$i\in F$, such that

i) $\rho_{i,j}=\rho_{j,i}$, $i\neq j$,

ii) $\rho_i$, $\rho_{i,j}$ are non-negative reals.
 \ali
The terms $(\rho_{i,j})$ are called the conductances, and
the terms $(\rho_i)$ are the dissipative terms.
We say that the electrical network is
 irreducible when the graph defined by the strictly positive
 $\rho_{i,j}$ is connected. We say that $\rho$ is conservative when
$\rho_i=0$ for all $i$.
 \ali
With $\rho$, we associate the element $Q_\rho$ in $\symF(\BR)$, by
$$
(Q_\rho)_{i,j}=\left\{
\begin{array}{l}
-\rho_{i,j}, \;\; i\neq j,
\\
\rho_{i}+\sum_{k\neq i} \rho_{i,k},\;\; i=j.
\end{array}
\right.
$$
The energy dissipated by the network, for the potential
$f:F\rightarrow \BR$, is given by the quadratic form:
$$
\eee^\rho(f,f)=\;<Q_\rho f,f>\;=\sum_{i\in F} f(i)^2 \rho_i
+\demi\sum_{i\neq j} \rho_{i,j} (f(i)-f(j))^2,
$$
where $<\cdot ,\cdot>$ is the usual scalar product on $\BR^F$ (the
bilinear form $\eee^\rho(f,h)$ on $\BR^F\times \BR^F$ is defined
by polarization).
 \ali
  The electrical current associated with a potential $f$ is
the element $I^\rho_f$ of the dual space $(\BR^F)^*$ of
$\BR^F$, defined by
$$
I^\rho_f(h)= \eee^\rho (f,h), \;\;\; \forall h\in \BR^F.
$$
Of course, if $(e_i)$ is the canonical basis of $\BR^F$, and
$(e_i^*)$ the dual basis, then we have
$$
I_f^\rho =\sum_{i\in F} \left( \sum_{j\in F} (Q_\rho)_{i,j} f(j)
\right) e_i^*.
$$
We denote by $\ddd_F\subset \symF(\BR)$ the positive cone of real
symmetric operators of the type $Q_\rho$, for
$\rho=((\rho_{i,j}),(\rho_i))$ a dissipative electrical network.
We denote by $\ddd_F^0$ the subcone of $\ddd_F$ consisting of
elements of the type $Q_\rho$ for conservative electrical
networks.
 \ali
{\it Probabilistic interpretation.} When $Q_\rho\in \ddd_F$, the
bilinear form $\eee^\rho(\cdot, \cdot)$ is a Dirichlet form on the
set $F$ (cf. \cite{Fukushima1}). If $b$ is a positive measure on
the set $F$, the symmetric operator $H_{\rho,b}$, on $\BR^F$,
defined by
$$
<Q_\rho f,h >= -\int H_{\rho,b} f \cdot h db, \;\;\; \forall
f,h\in \BR^F,
$$
is the infinitesimal generator of a discrete Markov process
behaving as follows: the process waits an exponential time
of parameter
${1\over b(\{i_0\})}(\rho_{i_0}+\sum_{j\neq i_0} \rho_{i_0,j})$ at
a point $i_0$ and then is killed with probability
${\rho_{i_0}\over \rho_{i_0}+\sum_{j\neq i_0} \rho_{i_0,j}}$ or
jumps to a point $j_0\neq i_0$ with probability
${\rho_{i_0,j_0}\over \rho_{i_0}+\sum_{j\neq i_0} \rho_{i_0,j}}$.
The set $\ddd_F^0$ corresponds to conservative Dirichlet forms. In
this case there is no killing part.

\subsection{Lagrangian compactification}
We set $E=\BC^F$, and denote by $E^*=(\BC^F)^*$ the dual space. We
denote by $(e_i)_{i\in F}$ the canonical basis of $E$ and by
$(e^*_i)_{i\in F}$ the dual basis. Let us set $V_F=E\oplus E^*$
(sometimes we write $V_K$ or simply $V$ when no ambiguity is
possible), and denote by $(,)$ the canonical symmetric bilinear
form, and by $<,>$, the canonical Hermitian scalar product on $V$,
given by
$$
(X,Y)=\sum_{i=1}^{2K} X_i Y_i, \;\;\;
<X,Y>=\sum_{i=1}^{2N}\overline X_i Y_i,
$$
where $X_i$, $Y_i$ are the coordinates of $X$, $Y$ in the basis
$((e_i),(e^*_i))$. When we consider the real part, we write
$E_\BR$ for $\BR^F$ and $E_\BR^*=(\BR^F)^*$. Let $\w$ be the
canonical symplectic bilinear form on $V_F\times V_F$ given by
$$
\w((x,\ksi),(x',\ksi'))=\ksi'(x)-\ksi(x'),
$$
for all $(x,\ksi)$ and $(x',\ksi')$ in $V_F\simeq E\times E^*$.
We denote by $\perp_\w$ the orthogonality relation for the
bilinear form $\w$. For any subspace $L\subset V$, we denote by
$L^o$ the orthogonal subspace of $L$ for the bilinear form $\w$.

Let $J$ be the antisymmetric operator on $V=E\oplus E^*$ defined
by block by
$$
J=\left(
\begin{array}{cc}
0& -\Id
\\
\Id& 0
\end{array}
\right).
$$
Clearly, we have
\begin{eqnarray}\label{f.1.JX}
\w(X,Y)= (JX,Y), \;\;\; \w(X, Y)=<J\overline{X},Y>.
\end{eqnarray}
We denote by $\perps$, the orthogonality relation for $<,>$. For a
subspace $L\subset V$, we denote by $L^\perp$ its orthogonal
complement, for the Hermitian scalar product $<,>$. It is clear
with these notations, that for any subspace $L\subset V$, we have
$$
L^o=\overline{JL^\perp}, \;\;\; L^\perp =\overline{JL^o}.
$$
(N.B.: Here and in the following, $\overline{JL^\perp}$ and
$\overline{JL^o}$ represent the complex conjugation of the linear
spaces $JL^\perp$ and $JL^o$.)
 \ali
 Indeed, $\overline{JL^\perp}$ has the right dimension and
$\overline{JL^\perp}\perp_{\w} L$, using formula (\ref{f.1.JX}).
\begin{defin} A vector subspace $L\subset V$ is isotropic (resp.
coisotropic) if $L\subset L^{o}$ (resp. $L^o \subset L$). We say
that $L$ is Lagrangian if $L^o =L$. Lagrangian subspaces have
dimension equal to $\dim E=\vert F\vert=K$.
\end{defin}
If $L$ is Lagrangian, then clearly, $L^\perp =\overline{JL}$.

We denote by $\lll_F$, resp. $\lll_{F,\BR}$, the set of Lagrangian
subspaces of $V$, resp. of real Lagrangian subspaces of
$V_\BR=E_\BR\oplus E^*_\BR$ (sometimes we write $\lll_K$ or
$\lll_V$ instead). The set  $\lll_F$ has the structure of a smooth
subvariety of $G_\BC(K,2K)$, the complex Grassmannian of
$K$-dimensional subspaces of $\BC^{2K}$ (indeed, $\lll_F$ is
isomorphic to $Sp(K, \BC)/ P_K$ where $Sp(K,\BC)$ is the
symplectic linear group and $P_K$ a maximal parabolic subgroup,
cf. \cite{Sabot5}, appendix E). The tangent space at a point is
isomorphic to $\sym_K(\BC)$ and we now give an explicit local
parameterization of $\lll_F$. Let $L$ be a Lagrangian subspace of
$V$. Let $(v_1,\ldots ,v_K)$ be an orthonormal basis of $L$ and
set $(v_1^*,\ldots ,v_K^*)=\overline{J(v_1,\ldots ,v_K)}$, which
is an orthonormal basis of $L^\perp =\overline{JL}$. For $Q$ in
$\sym_K(\BC)$, we set
 $$v_i^Q=v_i+\sum_{j=1}^K Q_{i,j} v_j^*.$$
The subspace generated by the family $\{v_i^Q\}_{i=1}^K$, is
Lagrangian. The map
\begin{eqnarray}
\nonu \sym_K(\BC) &\rightarrow & \BL_F
\\
\label{f.2.12} Q&\mapsto & \hbox{Vect}\{ v_i^Q \}_{i=1}^K
\end{eqnarray}
defines a local set of coordinates. Indeed, it is easy to check
that any Lagrangian subspace in a neighborhood of $L$ can be
represented in such a form.

 Considering this local
parameterization at the point $E\oplus 0$, with the basis
$(v_1,\ldots ,v_K)=(e_1,\ldots ,e_K)$, gives a natural embedding
of $\sym_F(\BC)$ in $\lll$. More precisely, with any point $Q$ in
$\sym_F(\BC)$ we associate the subspace $L_Q$ in $\lll_F$ given by
\begin{eqnarray} \label{f.1.2}
L_Q=\vect\{e_i+\sum_{j} Q_{i,j} e^*_j\}_{i\in F},
\end{eqnarray}
where we recall that $(e_1, \ldots ,e_K, e_1^*,\ldots ,e_K^*)$ is
the canonical basis of $V=E\oplus E^*$. With this embedding, the
set $\lll_F\setminus \symF(\BC)$ is exactly the set
$$
\lll_F\setminus \symF(\BC)=\{ L\in \lll_F,\;\; L\cap(0\oplus
E^*)\neq \{0\}\}
$$
which is an analytic subvariety of codimension 1 in $\lll_F$.
Hence, $\lll_F$ is a compactification of $\symF(\BC)$.
If $\rho$ is an electrical network, we see that
$$
L_{Q_\rho}=\{f+I_f^\rho, \;\; f\in \BR^F\},
$$
where $I_f^\rho$ is the current defined in section 1.1.
\begin{remark}
The set of Dirichlet forms $\ddd_F$ is thus a subset of $\lll_{F,\BR}$
and the closure of $\ddd_F$ in $\lll_F$ gives a compactification
of $\ddd_F$, which is described in theorem 5 of \cite{Colin1}.
\end{remark}
\subsection{Symplectic reduction}
Let $W$ be a coisotropic subspace of $V$, with dimension $K+p$,
$p\ge 0$ (the dimension of $W$ is necessarily greater or equal to
$K$ since $W^o\subset W$ and $\dim W+\dim W^o =\dim V=2K$). The
quotient space $W/W^o$ has dimension $2p$ and the symplectic form
$\w$ on $V$ naturally induces a symplectic form $\w_{W/W^o}$ on
$W/W^o$. Indeed, if $X$, $Y$ are in $W/W^o$, we define
\begin{eqnarray}\label{f.new.0}
\w_{W/W^o}(X,Y)= \w(\tilde X,\tilde Y),
\end{eqnarray}
where $\tilde X$ and $\tilde Y$ are any representatives in $W$ of
the quotient class $X$ and $Y$. (The right-hand side does not
depend on the choice of $\tilde X$ and $\tilde Y$ since
$W^o\perp_\w W$). The bilinear form $\w_{W/W^o}$ is antisymmetric
and non-degenerate (indeed, the $\w_{W/W^o}$-orthogonal of $W/W^o$
is $\{0\}$, by construction), thus it is a symplectic form.

For any subspace $L$ of $W$, we set
$$
t_W(L)= (L\cap W)/W^o,
$$
(more precisely, we mean that $t_W(L)$ is the subspace equal to
the projection of $L\cap W$ to $W/W^o$). We have
$$
\left(t_W(L)\right)^o =t_W(L^o),
$$
where on the left-hand side, the term $\left(t_W(L)\right)^o$
stands for the $\w_{W/W^o}$-orthogonal of $t_W(L)$. Indeed, we
have
$$
(L\cap W)^o \cap W=(L^o+W^o)\cap W= L^o\cap W+W^o,
$$
and
\begin{eqnarray*}
\left(t_W(L)\right)^o &=& \left( L\cap W/W^o\right)^o
\\
&=& \left( (L\cap W)^o \cap W\right) /W^o
\\
&=& \left( L^o\cap W + W^o\right)/W^o \\
&=& (L^o\cap W)/W^o =t_W(L^o).
\end{eqnarray*}
It implies that if $L$ is isotropic (resp. coisotropic, resp.
Lagrangian), then $t_W(L)$ is an isotropic (resp. coisotropic,
resp. Lagrangian) subspace of $W/W^o$. In this text, $L$ will
always be Lagrangian, and we will consider the  symplectic
reductions $t_W$ as a map from $\lll_F$ to $\lll_{W/W^o}$ (where
$\lll_{W/W^o}$ is the Grassmannian of Lagrangian subspaces of
$W/W^o$).
 \ali \ali
{\it Composition of symplectic reductions.}
 \ali
If $W'$ is a  subspace of $W/W^o$, we denote by $W'+W^0$ the
preimage of $W'$ by the canonical projection $W\mapsto W/W^0$.
Remark that we have
$$
(W'+W^0)^0=((W')^o+W^0).
$$
(Indeed, we have $(W'+W^0)^0\subset W$, and  the previous formula
comes from the definition of $\w_{W/W^0}$, cf. (\ref{f.new.0}).)
In particular, if $W'$ is $\w_{W/W^0}$-coisotropic, then $W'+W^0$
is $\w$-coisotropic and we have the following formula
\begin{equation}
\label{composition}
t_{W'}\circ t_W =t_{W'+W^0}.
\end{equation}
Indeed, if $L$ is a subspace of $V$, then we have
\begin{eqnarray*}
\left( \left( L\cap W/W^0\right)\cap W'\right)/(W')^0 &=&
\left(\left( L\cap (W'+W^0)\right)/W^0\right)/(W')^0
\\
&=& L\cap (W'+W^0)/((W')^0+W^0)
\\
&=& t_{W'+W^0}(L).
\end{eqnarray*}

\subsection{Pl\"ucker embedding}

As a subvariety of the Grassmannian $G_\BC(K ,2K)$, $\lll_F$ can
be embedded in a projective space by the Pl\"ucker embedding. It
will often be useful to consider this embedding since it gives a
set of homogeneous coordinates to represent the points of
$\lll_F$. In particular, it will be useful in order to represent
some rational maps on $\lll_F$ by homogeneous polynomial maps
through the Pl\"ucker embedding.

 We consider the exterior product
$$
\bigwedge^K (E\oplus E^*)\simeq \bigoplus_{k=0}^K
(\bigwedge^k E)\otimes
(\bigwedge^{K-k} E^*),
$$
and denote by $\ppp(\bigwedge^K (E\oplus E^*))$ the associated
projective space and by $\pi:\bigwedge^K (E\oplus E^*)\rightarrow
\ppp(\bigwedge^K (E\oplus E^*))$ the canonical projection.
Classically, the manifold $\lll_F$ can be embedded in the
projective space $\ppp (\wedge^K(E\oplus E^*))$ by the Pl\"ucker
embedding
\begin{eqnarray}
\nonu \lll &\rightarrow & \ppp (\wedge^K (E\oplus E^*))
\\
\label{f.1.3} L= \hbox{Vect}\{ x_1,\ldots ,x_K\} &\mapsto & \pi (
x_1\wedge \cdots \wedge x_K).
\end{eqnarray}

{\it Grassmann algebra.} When there is a canonical splitting of
the space $V$, as is the case here with $V=E\oplus E^*$, then it
is sometimes easier to represent this embedding a bit differently.
Let $(\oeta_i)_{i\in F}$ and $(\eta_i)_{i\in F}$ be two sets of
variables, and consider the Grasmann algebra generated by these
variables, i.e. the $\BC$-algebra generated by $(\oeta_i)_{i\in
F}$ and $(\eta_i)_{i\in F}$ with the anticommuting relations
$$
\eta_i \eta_j =-\eta_j \eta_j, \;\; \eta_i \oeta_j =-\oeta_j
\eta_j, \;\; \oeta_i \oeta_j =-\oeta_j \oeta_i.
$$
We denote by $\aaa$ the subalgebra generated by the monomials
containing the same number of variables $\oeta$ and $\eta$
(clearly, $\aaa$ is isomorphic to $\oplus_{k=0}^K \bigwedge^k E
\otimes \bigwedge^k E$). A canonical basis of $\aaa$ is
$$
(1, \oeta_{i_1} \cdots  \oeta_{i_k} \eta_{j_1} \cdots
\eta_{j_{k}},\; i_1<\cdots <i_k,\; j_1<\cdots <j_{k},\; 1\le k\le
K ).
$$
We endow $\aaa$ with $<\cdot ,\cdot >$, the Hermitian scalar
product which makes this basis an orthonormal basis (with the
convention that it is linear on the right and sesquilinear on the
left).

If $Q$ is a $K\times K$ matrix, then we denote $\oeta Q\eta$ the
element of $\aaa$:
$$\oeta Q\eta =\sum_{i,j\in F} Q_{i,j} \oeta_i \eta_j.$$
We will be particularly interested in terms of the type
\begin{eqnarray}
\nonu \exp (\oeta Q \eta )&=& \sum_{k=0}^K {1\over k!}
\left(\sum_{i,j} Q_{i,j} \oeta_i \eta_j \right)^k
\\ \nonumber
&=& \sum_{k=0}^K \sum_{{i_1<\cdots <i_k \atop j_1<\cdots <j_k}}
\det \left( (Q)_{{i_1,\ldots ,i_k \atop j_1,\ldots ,j_k}} \right)
\oeta_{i_1} \eta_{j_1}\cdots \oeta_{i_k} \eta_{j_k},
\end{eqnarray}
where $(Q)_{{i_1,\ldots ,i_k \atop j_1,\ldots ,j_k}}$ is the
$k\times k$ matrix obtained from $Q$ by keeping only the lines
$i_1,\ldots ,i_k$ and the columns $j_1,\ldots ,j_k$.

The algebra $\aaa$ is clearly isomorphic to $\bigwedge^K (E\oplus
E^*)$ by the isomorphism $\tau:\aaa\rightarrow \bigwedge^K(E\oplus
E^*)$ given on the elements of the basis by
\begin{eqnarray}
\label{f.1.5}
\tau(\oeta_{i_1}\eta_{j_1} \cdots \oeta_{i_k}\eta_{j_k})=
e_1\wedge \cdots \wedge \build{}{\check{e^*}_{j_1}}{i_1}\wedge
\cdots \wedge \build{}{\check{e^*}_{j_k}}{i_k}\wedge \cdots \wedge e_K,
\end{eqnarray}
for all $k\le K$, $i_1< \cdots < i_k$, $j_1<\cdots <j_k$, and where
we write $\cdots \wedge \build{}{\check{e^*}_{j}}{i}\wedge \cdots$ for
the element obtained by replacing the term $e_i$ by $e_j^*$ in the
monomial $e_1\wedge \cdots \wedge e_K$.

It is clear from formula (\ref{f.1.5}), that, for $Q$ in
$\symF(\BC)$, we have
$$\tau(\exp\oeta Q\eta)=
(e_1+\sum_{j=1}^K Q_{1,j}
e_j^*)\wedge \cdots \wedge (e_K+\sum_{j=1}^K Q_{K,j}
e_j^*).
$$
Hence, through the isomorphism $\tau$, the subset $\symF(\BC)$ is
embedded in the projective space $\ppp(\aaa)$ by
\begin{eqnarray*}
\symF &\hookrightarrow& \ppp(\aaa)
\\
Q &\mapsto &\pi(\exp(\oeta Q\eta)).
\end{eqnarray*}

\section{Trace map, gluing, and symplectic reduction}
\subsection{Trace map}
Let $\bF$ be a non-empty subset of $F$ ($\pF$ plays the role of  a
boundary set for $F$). Here, we describe an operator which plays a
key role in the analysis of self-similar Schr\"odinger operators
on pcf self-similar sets. Following the terminology of Dirichlet
forms we called this operator the ``trace operator" in
\cite{Sabot5}, but it bears several different names and appears in
several different fields of mathematics: for example, it is called
Neumann to Dirichlet operator in the theory of differential
operators or ``answer of a network" in the context of electrical
networks (cf. \cite{Colin1}), but also trace operator in the
theory of Dirichlet forms (cf. \cite{Fukushima1}, part 6) and
Schubert's complement in linear algebra (cf. for example
\cite{Metz2}, \cite{Carlson}).

We first describe this map on the set of real, symmetric,
non-negative $K\times K$ matrices. Let $Q$ be in $\symF(\BR)$ and
non-negative. We denote by $\bQ$ the real symmetric operator on
$\BR^\bF$ defined by the following variational problem
$$
<\bQ f,f>\; =\inf_{{g\in\BR^F, \atop g_{|\bF}=f}} <Qg,g>, \;\;\;
\forall f\in \BR^{\bF},
$$
where $<\cdot ,\cdot >$ denotes the usual scalar product
respectively on $\BR^\bF$ and $\BR^F$. We take from \cite{Sabot5},
proposition 2.1, the following simple properties.
\begin{propos} \label{p.2.1}
The map $Q\mapsto \bQ$ has a rational extension to $\symF(\BC)$
given by
\begin{eqnarray}
\label{f.2.1} Q_{\bF}=Q_{|\bF}-B(Q_{|F\setminus \bF})^{-1}B^t,
\end{eqnarray}
when $Q$ has the following block decomposition on $\bF$ and
$F\setminus \bF$:
\begin{eqnarray*}
Q= \left(
\begin{array}{cc}
Q_{|\bF}&B\\
B^t& Q_{|F\setminus \bF}
\end{array}
\right).
\end{eqnarray*}
Therefore the map $Q\mapsto Q_{\bF}$ is rational in the
coefficients of $Q$ with poles included in the set $\{\det
(Q_{|F\setminus \bF})=0\}$.

ii) If $\det(Q_{|F\setminus \bF})\neq 0$, then for any function
$f$ in $\BC^{\bF}$, we denote by $Hf$, the function of $\BC^F$
given by
 \beqn \left\{
\begin{array}{l}
Hf=f \;\;\hbox{on $\bF$,}
\\
Hf=-(Q_{|F\setminus \bF})^{-1}B^t f\;\;\hbox{on $F\setminus \bF$.}
\end{array}
\right.
 \eeqn
We call $Hf$ the harmonic extension of $f$ with respect to $Q$ and
we have $Q_{\bF}(f)=(Q(Hf))_{|\bF}$.
\end{propos}
\begin{remark}
The trace map sends the cone of Dirichlet forms $\ddd_F$ (resp.
$\ddd^0_F$) to the cone of Dirichlet forms $\ddd_{\pF}$ (resp.
$\ddd_\pF^0$). It means that if $\rho$ is a dissipative (resp.
conservative) electrical network, then there exists a dissipative
(resp. conservative) electrical network $\rho_\pF$ on $\pF$ such
that $(Q_\rho)_\pF= Q_{\rho^\pF}$ (cf. for example \cite{Sabot1},
proposition 1.9). This operation has a probabilistic
interpretation in terms of the underlying Markov process, cf.
\cite{Fukushima1}, theorem 6.2.1.
\end{remark}
 Let us now describe the extension of this map to $\lll_F$.
As shown in \cite{Colin1}, this corresponds to a symplectic
reduction. We set
$$
W_\pF= \BC^F\oplus (\BC^\bF)^* \subset E\oplus E^*.
$$
(In this section we sometimes
simply write $W$ to simplify notations).
 The $\w$-orthogonal subspace of $W$ is
$W^o=\BC^{F\setminus \bF}\oplus 0$.
 Remark first
that $W/W^o$ can be identified with $V_\pF=\BC^{\bF}\oplus
(\BC^\bF)^*$, and that the restriction of $\w$ to $W$ naturally
induces the canonical symplectic form $w_{\partial F}$ on
$W/W^o\sim \BC^\bF\oplus (\BC^\bF)^*$ (cf. section 1.3).
 \ali
If $L$ is a Lagrangian subspace of $E\oplus E^*$, then we set
\begin{eqnarray}\label{f.2.4}
t_{F\rightarrow \bF} (L)=t_{W_\pF}(L)=(L\cap W)/W^o\in W/W^o.
\end{eqnarray}
We know, from section 1.3, that $t_{F\rightarrow \bF} (L)$ is a
Lagrangian subspace of $V_\pF$.  We have the following proposition
(cf. \cite{Colin1}, section 5.1).
\begin{propos}\label{p.2.2} The map $t_{F\rightarrow
\bF}$ coincides with the map $Q\mapsto Q_\bF$ on the set
$\symF(\BC)\setminus\{\det Q_{|F\setminus \bF}=0\}$, through the
embedding of $\symF(\BC)$ and $\sym_\bF(\BC)$ respectively in
$\lll_F$ and $\lll_\bF$ (as described in section 1.2). Otherwise
stated, this means that
$$
t_{F\rightarrow \pF}(L_Q)=L_{Q_\pF},
$$
on the set $\{\det(Q_{|F\setminus \pF})\neq 0\}$.
\end{propos}
\begin{remark}
The  map $t_{F\rightarrow \bF}$ is not everywhere smooth. In
section 3, we describe its discontinuities.
\end{remark}
Proof: Let $p=\vert \pF\vert$. For $x$ in $\pF$ we set
$g_x=H(e_x)$. The vector $g_x$ can be written $\sum_{y\in F} c^x_y
e_y$ and we set
$$
g_x^*=\sum_{y\in F} c^x_y\sum_{z\in F} Q_{y,z} e^*_z.
$$
By definition, $g_x+g_x^*$ is in $L_Q\cap W$ for all $x$ in $\pF$
and
$$
p_{W/W^o}(\vect\{g_x+g_x^*\}_{x\in \pF})
$$
has dimension $p$. This immediately implies that
$$
t_{F\rightarrow \pF}(L_Q)=\vect\{e_x+(g_x^*)_{|\pF}\}_{x\in \pF}
$$
and hence that $t_{F\rightarrow
\pF}(L_Q)=L_{Q_\pF}$.$\diamondsuit$
\begin{remark}
When the Lagrangian subspace is of the type $L_Q$, it is easy to
describe $t_{F\rightarrow \pF}(L_Q)$. Indeed, consider the space
of solutions $f:F\rightarrow \BC$ of
\begin{equation}
\label{equa-schro-Q}
(Qf)_{|F\setminus \pF}=0.
\end{equation}
For any  solution $f$ of the previous equation, the current
$I_f^Q$, defined as the element of $(\BC^F)^*$ such that
$I_f^Q(h)=<Qf,h>$ for all $h\in \BC^F$, is supported by $\pF$ and
hence is an element of $(\BC^\pF)^*$. Then
\begin{equation}
\label{LQ-pF}
 t_{F\rightarrow \pF}(L_Q)=\{f_{|\pF}+I_f^Q, \;\hbox{ for all $f$ solution
of (\ref{equa-schro-Q})}\}.
\end{equation}
This expression is interesting when, for an electrical network
$\rho$ and a positive measure $b$ on $F$, we consider
$Q=Q_\rho+\lambda I_b$, where
$I_b$ is the diagonal matrix with diagonal terms $(I_b)_{x,x}=b(\{x\})$.
In this case the solutions of (\ref{equa-schro-Q}) are the solutions of a
discrete Schr\"odinger equation
$$
\left((H_{\rho,b}-\lambda )f\right)_{|F\setminus \pF}=0.
$$
(N.B.: $H_{\rho,b}$ is defined in section 1.1.) In particular, if
$f$ is an eigenfunction of $H_{\rho,b}$ with eigenvalue $\lambda$,
which does not vanish on the boundary (i.e. such that
$f_{|\pF}\neq 0$), then $I_f^Q=0$ and  $t_{F\rightarrow \pF}(L_Q)$
 intersects the Lagrangian subspace
$\BC^\pF\oplus 0$ non-trivially. Similarly, the intersection
\begin{equation}
\label{derniere}
t_{F\rightarrow \pF}(L_Q)\cap(0\oplus (\BC^\pF)^*)
\end{equation}
is related to the eigenfunctions with
Dirichlet boundary conditions. This will play an important role in
relation with the renormalization map we introduce in section 4.
Remark also that $t_{F\rightarrow \pF}(L_Q)$ is in $\sym_\pF$ if and
only if the intersection (\ref{derniere}) is $\{0\}$. This is true
 if $\ker (Q_{|F\setminus \pF})=\{0\}$, and this is coherent with
the fact that the set of singularities of the trace map $Q\mapsto
Q_\pF$ is included in the set $\{\det(Q_{|F\setminus \pF})=0\}$.
\end{remark}

%{\it Probabilistic interpretation}
%It is easy to see that $\bQ$ is well defined and that the subsets
%$\ddd_F$ and $\ddd^0$ are left invariant by $Q\mapsto \bQ$ (cf.
%proposition 2 of \cite{Colin1} or \cite{Sabot1}, ?).

\subsection{Gluing}
Suppose now that $\rrr$ is an equivalence relation on $F$. We
denote by $\pi:F\rightarrow \rF$ the canonical surjection and by
$s:\BC^\rF\rightarrow \BC^F$ the linear map given by:
$$
s(f)=f\circ \pi, \;\;\; \forall f \in \BC^\rF.
$$
We denote by $s^t:\BC^F\rightarrow \BC^\rF$ the transposed map and by $s^*:
(\BC^F)^*\rightarrow (\BC^\rF)^*$ the dual map given by
\begin{eqnarray*}
s^t(e_x)= e_{\pi(x)}, \;\;  s^*(e_x^*)=e^*_{\pi(x)},\;\; \forall
x\in F,
\end{eqnarray*}
 where we recall that $(e_x)_{x\in F}$ and $(e_x^*)_{x\in F}$
 (resp. $(e_x)_{x\in \rF}$ and $(e_x^*)_{x\in \rF}$) are the
 canonical basis of $\BC^F$ and $(\BC^F)^*$ (resp. $\BC^\rF$ and
 $(\BC^\rF)^*$).
 If
$Q$ is a symmetric operator on $\BC^F$, it is natural to define
the linear operator $Q_\rF$ on $\BC^\rF$ by
$$
Q_\rF =s^t\circ Q\circ s.
$$
It is clear that if $Q$ is in $\ddd_F$ and associated with an
electrical network $\rho$, then $Q_\rF$ is in $\ddd_\rF$ and
associated with an electrical network $\rho^\rF$ given by
$$
\rho^\rF_{x,y}=\sum_{{x',y',  \; s.t.  \atop \pi(x')=x,
\pi(y')=y}} \rho_{x',y'}, \;\;\; \rho_x^\rF=\sum_{{x', \; s.t.
\atop \pi(x')=x}}\rho_{x'}.
$$

As previously, the extension of this map to $\lll_F$ is a
symplectic reduction. Indeed, let us consider the subspace
$W_{F/\rrr}\subset V_F$ (we sometimes simply write $W$ in this
section):
$$
W_\rF=\im(s)\oplus (\BC^F)^*=\BC^\rF\oplus (\BC^F)^*,
$$
where we considered $\BC^\rF$ as the subset $\{f\in \BC^F, \;
f(y)=f(y') \hbox{ if } \pi(y)=\pi(y')\}$ of $\BC^F$. We have,
\begin{eqnarray*}
W^o&=& 0\oplus (\ker s^*) \\
&=& \{(0,\ksi),\; \ksi\in (\BC^F)^* \hbox{ s. t. } \sum_{y, \;
\pi(y)=x} \ksi(e_y)=0, \; \forall x\in F\}.
\end{eqnarray*}
Clearly, $W/W^o$ can be identified with $\BC^\rF\oplus
(\BC^\rF)^*$ and the restriction $\w_{|W}$ induces the canonical
symplectic form $\w_{\rF}$ on $W/W^o$. We define the map
$t_{F\rightarrow F/\rrr}: \lll_F\rightarrow \lll_\rF$ by
$$
t_{F\rightarrow F/\rrr}(L)=t_{W_\rF}(L)=L\cap W/W^o.
$$
\begin{propos} \label{p.2.3}
The map $t_{F\rightarrow \rF}$ coincides with the map $Q\mapsto
Q_\rF$ on the set $\sym_F(\BC)$, i.e.
$$
t_{F\rightarrow \rF}(L_Q)=L_{Q_{\rF}}.
$$
\end{propos}
Proof: It is simple and left to the reader.$\diamondsuit$

\section{Properties of the symplectic reduction}
\subsection{Singularities of the symplectic reduction}
The trace map and the gluing map correspond to symplectic
reductions. The symplectic reduction is not everywhere continuous,
and in \cite{Sabot5}, the singularities play an important role in
the understanding of the spectral properties of the operator. In
this section, we consider the symplectic reduction from the
algebraic point of view, as a rational map, and determine
explicitly its indeterminacy points and its blow-up. We also
generalize proposition 2.3 of \cite{Sabot5}, which describes the
zeros of the corresponding map defined through the Pl\"ucker
embedding.

Let us first recall the definition of a rational map between
algebraic varieties (cf. for example \cite{GriffithsH}, pp.
490-493). Let $X$ and $Y$ be two algebraic varieties. We  denote
by $\pi_1\; \hbox{(resp. $\pi_2$)} : X\times Y \rightarrow X
\;\hbox{(resp. $Y$)}$ the two canonical projections. A rational
map $g$ from $X$ to $Y$ is defined by its graph
$$
\Gamma_g\subset X\times Y,
$$
when $\Gamma_g$ is an irreducible algebraic subvariety of $X\times
Y$ such that for all $x$ in the complement of a non-trivial
analytic subset of $X$, $\pi_1^{-1}(\{x\})\cap \Gamma_g$ is a
singleton. The subset $I\subset X$ where $\pi_1^{-1}(\{x\})\cap
\Gamma_g$ is not a singleton is called the set of indeterminacy
points of $g$. It is an analytic subset of codimension (strictly)
bigger than 1. (This comes from the fact that the graph $\Gamma_g$
is assumed to be irreducible.) The image of a point $x$ in $X$ is
defined by
$$
g(x)=\pi_2(\pi_1^{-1}(\{x\})).
$$
Hence, $g(x)$ is a single point for $x$ in $X\setminus I$.

Let $f:X\rightarrow Y$ and $g:Y\rightarrow Z$ be two rational maps
with indeterminacy sets $I_f$ and $I_g$, and such that
$f(X\setminus I_f)$ is not included
in  $I_g$. We define the composition
$g\circ f$ as the rational map defined by its graph
$$
\Gamma_{g\circ f}= \hbox{closure}\left(\left\{ \left(x,g(f (x))\right), \;\; \;
x \in X\setminus I_f, \; f(x)\in Y\setminus
I_g\right\}\right).
$$
\begin{remark}
The graph $\Gamma_{g\circ f}$ is not necessarily equal to the
graph
$$
\Gamma_g\circ \Gamma_f=\{(x,z)\in X\times Z, \;\; \exists y\in
Y,\; \hbox{ s.t. }\; (x,y)\in\Gamma_f,\; (y,z)\in \Gamma_g\}.
$$
This equality is true only when $\Gamma_g\circ \Gamma_f$ is
irreducible. This plays an important role in relation with the
degrees of the iterates $g^n$, when $g$ is a rational map from $X$
to itself (cf. for example \cite{DFavre} in the 2-dimensionnal
case, or \cite{Favre1}).
\end{remark}

Let us recall that $V=\BC^K\oplus (\BC^K)^*$, that $\w$ is the
canonical symplectic form on $V$,  and that $\lll_V$ is the
Grassmannian of Lagrangian subspaces of $V$. Let $W$ be a
coisotropic subspace of $V$, with dimension $K+p$, and $W^o$ its
$\w$-orthogonal. We first claim that
\begin{propos}\label{p.3.1}
The symplectic reduction $t_W: \lll_V\rightarrow \lll_{W/W^o}$ is
analytic on $\lll_V\setminus \{L, \; L\cap W^o\neq\{0\}\}$ and can
be extended into a rational map $\tilde t_W$ given by the graph
$$
\Gamma_{\tilde t_W}=\{(L,L')\in \lll_V\times \lll_{W/W^o}, \; \;
\dim(t_W(L)\cap L')\ge p-\dim(L\cap W^o)\}.
$$
In particular, the set of indeterminacy points is $\{L\in \lll_V,
\;\; L\cap W^o\neq \{0\}\}$, and for $L$ in the set of
indeterminacy points,  $t_W(L)$ is a particular point of the set
$\tilde t_W(L)$.
\end{propos}
\begin{remark}
This means that $\Gamma_{\tilde t_W}$ is the closure of the graph
of $t_W$.
\end{remark}
\begin{remark}
Even if this question seems natural, we could not find
such a result anywhere in the litterature.
\end{remark}
Proof: We first recall that $\overline{JW}=(W^o)^\perp$. Thus,
there is a natural $<,>$-orthogonal decomposition
$$
W=W\cap\overline{JW}\oplus W^o.
$$
Hence, we can canonically identify $W/W^o$ with
$W\cap\overline{JW}$, and we do so in the following.

We remark now that, with this canonical identification,
$\Gamma_{\tilde t_W}$ can be written in the equivalent form
\begin{eqnarray}\label{f.3.3}
\Gamma_{\tilde t_W}=\{(L,L')\in \lll_V\times \lll_{W/W^o}, \; \;
\dim(L\cap (L'\oplus W^o))\ge p\}.
\end{eqnarray}
Indeed, denote by $p_{W/W^o}:W\rightarrow W/W^o$ the canonical
projection. We have
\begin{eqnarray*}
\dim((L'\oplus W^o)\cap(L\cap W))&=&
\dim(p_{W/W^o}^{-1}(L')\cap(L\cap W))
\\
&=& \dim(L\cap \ker(p_{W/W^o}))+\dim(L'\cap p_{W/W^o}(L\cap W))
\\
&=& \dim(L\cap W^o)+\dim(L'\cap t_W(L)).
\end{eqnarray*}
This immediately implies formula (\ref{f.3.3}).

The fact that $t_W$ is analytic in $\{L\in \lll_V, \; L\cap
W^o=\{0\}\}$ is easy. Indeed, when $L\cap W^o=\{0\}$, the vector
subspaces $L$ and $W$ are in generic position, thus the
application $L\rightarrow L\cap W$ is analytic from $G(K,V)$ to
$G(p,W)$, respectively the Grassmannian of $K$ dimensional
subspaces of $V$ and the Grassmannian of $p$-dimensional subspaces
of $W$. Then, the application $L\cap W \mapsto p_{W/W^o}(L\cap W)$
is analytic on the set where $(L\cap W)\cap\ker p_{W/W^o}=\{0\}$.

Thus, the only thing we have to prove is that $\Gamma_{\tilde
t_W}$, defined in proposition \ref{p.3.1}, is equal to the closure
$$
\overline{\{(L,t_W(L)), \; L\in \lll_V, \hbox{ s.t. $L\cap
W^o=\{0\}$}\}}.
$$
Using the representation (\ref{f.3.3}), we already know that this
closure is included in $\Gamma_{\tilde t_W}$. Indeed, the
dimension of the intersection of two subspaces is semi-continuous
from below.

We prove now that for any $(L,L')$ in $\Gamma_{\tilde t_W}$, such
that $\dim(L\cap W^o)=n_0>0$, we can find $L_\epsilon$ in a small
neighborhood of $L$ such that $L_\epsilon \cap W^o=\{0\}$ and
$t_W(L)=L'$. Let us first prove this for $L'=t_W(L)$. We remark
first that we have the following $<,>$-orthogonal decomposition
\begin{eqnarray*}
L= L\cap W^o \oplus (L\cap W)\cap (L\cap W^o)^\perp \oplus L\cap
(L\cap W)^\perp.
\end{eqnarray*}
Take some orthonormal basis $f_1, \ldots ,f_{n_0}$ of $L\cap W^o$,
$f_{n_0+1}, \ldots , f_{p+n_0}$ of $(L\cap W)\cap (L\cap
W^o)^\perp$, and $f_{n_0+p+1}, \ldots ,f_K$ of $L\cap (L\cap
W)^\perp$. Define now $(f^*_1, \ldots
,f_K^*)=\overline{J(f_1,\ldots ,f_K)}$, and $f_i^\epsilon$ by
$f_i^\epsilon=f_i+\epsilon f_i^*$ for $i\le n_0$, and
$f_i^\epsilon=f_i$ for $i\ge n_0+1$. Then it is clear (cf. section
1.2) that $L_\epsilon$, the vector space generated by the family
$(f_i^\epsilon)$, is Lagrangian and satisfy both $L\cap W^o=\{0\}$
and $t_W(L_\epsilon)=t_W(L)$.

Let us consider now any $(L,L')$  in $\Gamma_{\tilde t_W}$ such
that $\dim(L\cap W^o)=n_0>0$. We just have to prove that in any
small neighborhood of $L$, we can find $L_\epsilon$ such that
$t_W(L_\epsilon)=L'$ (indeed, by a small modification, we can have
the extra property $L_\epsilon \cap W^o=\{0\}$). We set
$$
\tilde L=L'\cap t_W(L).
$$
The subspace $\tilde L$ has dimension $n_1\ge p-n_0$, by
hypothesis. Define $f_1, \ldots ,f_K$ as before. We can always
suppose that $f_{n_0+1}, \ldots ,f_{n_0+n_1}$ are such that
$$
p_{W/W^o}(\vect\{f_{n_0+1}, \ldots ,f_{n_0+n_1}\})=\tilde L.
$$
 We set
$$
L_1'=L'\cap \tilde L^\perp.
$$
Take now any surjective linear map $T:L\cap W^o \mapsto L_1'$
(there exists such a map since $\dim(L\cap W^o)\ge \dim L_1'$).
For $\epsilon>0$ we set $f_i^\epsilon =f_i+\epsilon T(f_i)$ for
$i\le n_0$ and $f_i^\epsilon =f_i$ for $n_0+1\le i\le n_0+n_1$.
Then we define $K_\epsilon =\vect\{f_i^\epsilon, \; 1\le i\le
n_0+n_1\}$. It is clear that $p_{W/W^o}(K_\epsilon)=L'$. This
implies that $K_\epsilon$ is isotropic, since $L'$ is Lagrangian
in $W/W^o\sim W\cap \overline{JW}$. Hence, we can always construct
a symplectic transformation $S_\epsilon\in Sp(V)$, close to the
identity for $\epsilon$ small, such that $S_\epsilon
(K_0)=K_\epsilon$. Let us define $L_\epsilon= S_\epsilon(L)$,
which is a Lagrangian subspace of $V$, close to $L$ for small
$\epsilon$. By construction $K_\epsilon\subset L_\epsilon \cap W$,
hence $t_W(L_\epsilon)=L'$.$\diamondsuit$

\subsection{Linear lift by the Pl\"ucker embedding}

Remind that $\lll_V$ is embedded in the projective space
$\ppp(\wedge^K V)$ by the Pl\"uker embedding (cf. section 1.4).
Similarly, $\lll_{W/W^o}$ is embedded in $\ppp (\wedge^p W/W^o)$.
In this section we construct an explicit linear map $R_W:\wedge^K
V\rightarrow \wedge^p W/W^o$ which lifts the symplectic reduction
$t_W$.

We recall that $(W^o)^\perp =\overline{JW}$, $W^\perp =
\overline{JW^o}$ and that we have the orthogonal decomposition
$W=W\cap \overline{JW}\oplus W^o$, which gives a canonical
isomorphism between $W/W^o$ and $W\cap\overline{JW}$. We choose an
orthonormal basis $(g_1, \ldots ,g_{K-p})$ of $W^o$ and set
$(g_1^*,\ldots ,g_{K-p}^*)=\overline{J(g_1,\ldots ,g_{K-p})}$,
which gives an orthonormal basis of $W^\perp=\overline{JW^o}$.
 \ali
For $l\le K$ and $Y$ in $\bigwedge^l V$, we denote by
$i_Y:\bigwedge^K V\rightarrow \bigwedge^{K-l} V$, the interior
product defined as the linear map  on $\bigwedge^K V$ such that
$$
<Z,i_Y(X)>=<Y\wedge Z,X>,\;\;\; \forall X\in \bigwedge^K V, \;\;
\forall Z\in \bigwedge^{K-l}V,
$$
where $<,>$ is the Hermitian product induced by the canonical
Hermitian product on $V$.
 The interior product
$i_{g_1^*\wedge \cdots \wedge g_{K-p}^*}$ sends $\bigwedge^K V$ to
$\bigwedge^{p} W$ and we set
\begin{eqnarray*}
R_W:\bigwedge^K V &\rightarrow &
\bigwedge^p W/W^o
\\
X&\mapsto & (\wedge^{p}p_{W/W^o})\circ i_{g_1^*\wedge \cdots
\wedge g_{K-p}^*}(X),
\end{eqnarray*}
where $p_{W/W^o}:W\rightarrow W/W^o$ is the orthogonal projection
on $W\cap \overline{JW}\simeq W/W^o$.
\begin{remark}
The expression of $R_W$ is not very simple, but in the special
cases of the trace map and the gluing map,  the expression is
quite simple and natural (cf. the end of the section).
\end{remark}

\begin{remark}
 Up to a sign, the value of $R_W$ does not depend on the
particular choice of the orthonormal basis $(g_1, \ldots
,g_{K-p})$.
\end{remark}
Let us give a definition: if $f$ is a holomorphic function from a
domain $D\subset \BC^n$ to $\BC^m$, then we denote by
$\ord(f,x_0)$ the order of vanishing of $f$ at the point $x^0\in
D$, i.e. the maximal integer $p$ such that one can find an open
set $U$ containing $x_0$ and holomorphic functions $h_{i_1,\ldots
,i_p}$, $1\le i_1\le \ldots \le i_p\le n$ on $U$ such that
$$f=\sum_{i_1\le \cdots \le i_p}
(x_{i_1}-x_{i_1}^0)\cdots (x_{i_p}-x_{i_p}^0) h_{i_1,\ldots
,i_p}(x),\;\;\; \hbox{on $U$ .}
$$
Let us finally recall that we denote by $\pi$, both the canonical
projection $\pi: \bigwedge^K V \rightarrow \ppp ( \bigwedge^K V)$
and $\pi: \bigwedge^p W/W^o \rightarrow \ppp ( \bigwedge^p W/W^o)$.
\begin{propos} \label{p.3.RW}
i) If $L\in \lll_V$ is such that $L\cap W^o=\{0\}$, and $X_L\in
\bigwedge^K V\setminus \{0\}$ such that $\pi (X_L)=L$, then
$R_W(X_L)\neq 0$ and
$$
\pi(R_W(X_L))= t_W(L).
$$

ii) If $L\in \lll_V$ is such that $\dim(L\cap W^o)=n_0$, and $s$
is a local holomorphic section of $\pi$ on an open subset
$U\subset \lll_V$ containing $L$, then
$$
\ord (R_W\circ s, L)=n_0 .
$$
\end{propos}
\begin{remark}
 Otherwise stated, (i) means that the following diagram
commutes on the subset where all the maps are well-defined.
$$
\begin{CD}
{\pi^{-1}(\lll_V)} @>\mathrm{R_W}>> {\pi^{-1}(\lll_V)}\\
@VV{\mathrm{\pi}}V @VV{\mathrm{\pi}}V\\
{\lll_V}@>\mathrm{t_W}>> {\lll_V}
\end{CD}
$$
\end{remark}
\begin{remark}
This is a generalization of proposition 2.2, formula (30),
and proposition 2.3 of \cite{Sabot5}, to general symplectic
reductions. Remark also that in proposition 2.3 of \cite{Sabot5},
this result was proved only for real $Q_0$. Actually, this
restriction is not necessary, as shown in the previous
proposition. The proof we give here is also simpler than the proof
of \cite{Sabot5}.
\end{remark}
Proof: i) Let us consider $L$ in $\lll_V$, such that $L\cap
W^o=\{0\}$. The subspace $L$ can be decomposed orthogonally  in
$$
L= L\cap W\oplus (L\cap W)^\perp \cap L.
$$
We choose an orthonormal basis $(f_1,\ldots ,f_p)$ of $L\cap W$
and $(f_{p+1},\ldots ,f_K)$ of $(L\cap W)^\perp \cap L$. We
consider the orthogonal projection $p_{\overline{JW^o}}$ on
$\overline{JW^o}$. Clearly, $p_{\overline{JW^o}}$ is an
isomorphism from $(L\cap W)^\perp \cap L$ onto $\overline{JW^o}$,
since $\ker(p_{\overline{JW^o}})=W$ and since they have the same
dimension. Thus, we have
$$
i_{g_1^*\wedge \cdots \wedge g_{K-p}^*}(f_1\wedge \cdots \wedge
f_p\wedge \cdots \wedge f_K)= C f_1\wedge\cdots\wedge f_p,
$$
where $C=i_{g_1^*\wedge \cdots \wedge g_{K-p}^*}(f_{p+1}\wedge
\cdots \wedge f_K)$ is a  non-null complex scalar. It follows that
\begin{eqnarray*}
\pi\left( (\wedge^{p}p_{W/W^o})\circ i_{g_1^*\wedge \cdots \wedge
g_{K-p}^*}(f_1\wedge \cdots \wedge f_K)\right)&=&\pi\left(
p_{W/W^o}(f_1)\wedge \cdots\wedge p_{W/W^o}(f_p)\right)
\\
&=& t_W(L),
\end{eqnarray*}
which is exactly what we want.

ii) If now $\dim(L\cap W^o)=n_0>0$, we have the orthogonal
decomposition
$$
L=(L\cap W^o)\oplus (L\cap W^o)^\perp \cap (L\cap W)\oplus (L\cap
W)^\perp \cap L .
$$
We choose orthonormal bases $(f_1,\ldots ,f_{n_0})$ of $L\cap
W^o$, $(f_{n_0+1}, \ldots ,f_{n_0+p})$ of $(L\cap W^o)^\perp \cap
(L\cap W)$, and $(f_{n_0+p+1}, \ldots , f_K)$ of $(L\cap W)^\perp
\cap L $. As usual, we set $(f_1^*,\ldots ,f_K^*)=\overline{J(f_1,
\ldots ,f_K)}$. Recall that $(g_1, \ldots ,g_{K-p})$ is the
orthonormal basis we chose for $W^o$. We can as well suppose (up
to a change of sign in $R_W$) that $(f_1, \ldots
,f_{n_0})=(g_1,\ldots ,g_{n_0})$. For $i\ge n_0+p+1$, we can make
the orthogonal decomposition $f_i=f_i'+f_i'' $
 with $f_i'\in W$,
$f_i''\in \overline{JW^o}$. We have $f_i''\in \overline{J(L\cap
W^o)}^\perp\cap \overline{JW^o}$, since
$$0=\w(f_i,f_j)=\w(f_i,f_i'')=<f^*_i,f''_j>,$$
for $i\le n_0$ and $j\ge n_0+p+1$. Moreover,
$$
(g_1^*, \ldots , g^*_{n_0}, f''_{n_0+p+1},\ldots ,f_{n}'')
$$
form a basis of $\overline{JW^o}$. For $Q$ in $\sym_K(\BC)$, we
set $f_i^Q=f_i+\sum_{j=1}^n Q_{i,j}f_j^*$. In the neighborhood of
$L$, $\lll_V$ can be parametrized by $\sym_K(\BC)$, by
$$
Q\mapsto \vect\{f_i^Q\}_{i=1}^K.
$$
Then, we have
\begin{eqnarray}
\nonumber R_W(f_1^Q\wedge \cdots \wedge f_n^Q)
&=&C\det((Q_{i,j})_{i,j=1}^{n_0})\left( p_{W/W^o}f_{n_0+1}\wedge
\cdots
\wedge p_{W/W^o}f_{n_0+p} \right) \\
\label{f.3.RW} &&+(\hbox{terms of higher degree in $Q_{i,j}$}),
\end{eqnarray}
where $C=i_{g_1^*\wedge \cdots \wedge g_{K-p}^*}(f_1\wedge
\cdots\wedge f_{n_0}\wedge f_{n_0+p+1}''\wedge \cdots \wedge
f_K'')$ is a non-null complex scalar. This immediately implies ii)
of the proposition since $\bigwedge^p p_{W/W_0}(f_{n_0+1}\wedge
\cdots \wedge f_{n_0+p})$ is non null.$\diamondsuit$
 \ali\ali
{\it The corresponding map on the Grassmann algebra, in the case
of section 2.}

We come back to the situation of the trace map and the gluing map,
described in section 2. We use the Grassmann algebra $\aaa_F$ to
give the explicit expression of the map $R_W$, since the
expressions are simpler (and have an interpretation in terms of
antisymmetric integrals).

Let us first come back to the case of the trace map.
We denote by $\aaa_\bF$ the Grassmann algebra associated with the
set $\bF$, as in section 1.4. The algebra $\aaa_\bF$ corresponds
also to the subalgebra of $\aaa_F$ generated by the monomials
containing only the variables $\oeta_x$ and $\eta_x$, for $x$ in
$\bF$.

If $Y$ is in $\aaa$ we denote by $i_Y$ the interior product by
$Y$, i.e. the linear operator $i_Y\;:\;\aaa\rightarrow \aaa$
defined by
 \beq
  <Z,i_Y(X)>\;=\;<Y Z,X>, \;\;\; \forall
X,Z\in \aaa.
 \eeq
  In particular, remark that
$$i_{\Pi_{x\in F} \oeta_x\eta_x} (\exp \oeta Q\eta ))=\det Q.
$$
We define the linear operator
\begin{eqnarray}
R_{F\rightarrow \bF}\;:\; \aaa&\rightarrow & \aaa_{\bF}
\\
X&\mapsto & i_{\Pi_{x\in F\setminus \bF} \oeta_x \eta_x} (X).
\end{eqnarray}
\begin{remark}
The operator $R_{F\rightarrow \bF}$ is often presented as an
antisymmetric integral. More precisely, $R_{F\rightarrow \bF}(X)$
coincides with the antisymmetric integral of $X$ with respect to
$\Pi_{x\in F\setminus \bF} d\eta_xd\oeta_x$, i.e. $R_{F\rightarrow
\bF}(X) =\int X\Pi_{x\in F\setminus \bF} d\eta_x\oeta_x$, as
defined in \cite{Berezin} (cf. also \cite{Wang}).
\end{remark}
\begin{lem}
The operator $R_{F\rightarrow \pF}$ corresponds,
up to a sign, to the operator $R_{W_\pF}$ for the coisotropic subspace
$W_\pF$ defined in section 2.1. More precisely, it means that $R_{W_\pF}\circ\tau=\pm
\tau\circ R_{F\rightarrow \pF}$, where $\tau$ is the isomorphism defined in section
1.4
\end{lem}
Proof:
 We  can easily check this on
the elements of the canonical basis. Suppose that $\vert \pF\vert
=p\le K$, and that $\pF=\{K-p+1, \ldots ,K\}$. We have $(W_\pF)^o=
\BC^{F\setminus \pF}\oplus 0$. We take $(g_1,\ldots
,g_{K-p})=(e_1,\ldots ,e_{K-p})$, which is a basis of $(W_\pF)^o$.
We have $(g_1^*,\ldots  ,g_{K-p}^*)=(e_1^*,\ldots ,e_{K-p}^*)$.
Consider now an element of the basis of the type (with the
notations of section 1.4)
$$
e_1\wedge \cdots \wedge \build{}{\check{e^*}_{j_1}}{i_1}\wedge
\cdots \wedge \build{}{\check{e^*}_{j_k}}{i_k}\wedge \cdots \wedge
e_K.
$$
for $k\le K$, $i_1< \cdots < i_k$, $j_1<\cdots <j_k$.
It corresponds to $\oeta_{i_1}\eta_{j_1} \cdots \oeta_{i_k}\eta_{j_k}$
by the ismorphism $\tau$ defined in section 1.4.  It is clear that
$R_{W_\pF}$ is non null on this element
 if and only if $\{i_1,\ldots ,i_k\}\supset \{1,\ldots ,K-p\}$,
and $\{j_1,\ldots ,j_k\}\supset \{1,\ldots ,K-p\}$, i.e.
$R_{W_\pF}$ is non null on the elements of the
type
$$
(e_1^*\wedge \cdots\wedge e^*_{K-p})\wedge e_{K-p+1}\wedge \cdots
\wedge \build{}{\check{e^*}_{j_{K-p+1}}}{i_{K-p+1}}\wedge \cdots
\wedge \build{}{\check{e^*}_{j_k}}{i_k}\wedge \cdots \wedge e_K.
$$
The map $R_{W_\pF}$ applied to the previous element gives
$$
e_{K-p+1}\wedge \cdots \wedge \build{}{\check{e^*}_{j_{K-p+1}}}{i_{K-p+1}}\wedge
\cdots \wedge \build{}{\check{e^*}_{j_k}}{i_k}\wedge \cdots \wedge e_K,
$$
which corresponds to the element
 $\oeta_{i_{K-p+1}}\eta_{j_{K-p+1}} \cdots \oeta_{i_k}\eta_{j_k}$ in
$\aaa_\pF$. The latter is also equal to
$$R_{F\rightarrow \pF}(
\oeta_1\eta_1\cdots \oeta_{K-p}\eta_{K-p}\oeta_{i_{K-p+1}}\eta_{j_{K-p+1}}
 \cdots \oeta_{i_k}\eta_{j_k}).
$$
This is exactly the equality we need.$\diamondsuit$ \ali \ali It
means that the map $R_{F\rightarrow \pF}$ lifts the trace map
$Q\mapsto Q_\pF$ and the associated symplectic reduction
$t_{W_\pF}$. This was already proved in \cite{Sabot5}, where we
proved the following formula
\begin{eqnarray}\label{f.2.RFpF}
R_{F\rightarrow \partial F} (\exp\oeta Q\eta)= \det(Q_{|F\setminus
\bF}) \exp\oeta Q_\bF \eta.
\end{eqnarray}
Let us now remark that, in the case of the trace map, for a point of
the type $L_Q=\pi(\exp\oeta Q\eta)$, the set $L_Q\cap
W^o_\pF=\ker^\ND(Q)\oplus 0$, where
$$
\ker^\ND(Q)=\{f\in \BC^F, \; Qf=0 \hbox{ and } f_{|\pF}=0\}.
$$
Hence, the order of vanishing of $R_{F\rightarrow \pF}$ at $L_Q$
is equal to $\dim\ker^\ND(Q)$. (N.B.: we write $\ker^\ND$ for
``Neumann-Dirichlet kernel", in reference to the Neumann-Dirichlet
spectrum which plays an important role in section 4.)

Let us now consider the case of the ``gluing map". We denote by
$\aaa_\rF$ the Grassmann algebra associated with the set $\rF$, as
in section 1.4.
 The canonical surjection $\pi:F\rightarrow \rF$ naturally induces a
morphism of algebra $R_{F\rightarrow \rF}:\aaa_F\rightarrow
\aaa_\rF$ defined on generating variables by
$$
R_{F\rightarrow\rF}(\oeta_x)=\oeta_{\pi(x)}, \;\;\;
R_{F\rightarrow\rF}(\eta_x)=\eta_{\pi(x)}.
$$
\begin{lem}
The linear map $R_{F\rightarrow\rF}$
corresponds to the map $R_{W_\rF}$, up to a sign, for the
coisotropic subspace $W_\rF$  introduced in section 2.2. Otherwise stated
it means that
$R_{W_\rF}\circ\tau=\pm
\tau\circ R_{F\rightarrow \rF}$.
\end{lem}
Proof: it is simple, similar to the previous one, and left to the reader.$\diamondsuit$
 \ali
\ali
 Hence,
$R_{F\rightarrow\rF}$ lifts the ``gluing map" $Q\mapsto Q_\rF$ to
$\aaa_F$. It is actually very easy to check directly this last
point since we have the following trivial formula:
\begin{equation}
\label{RFrF}
R_{F\rightarrow \rF}(\exp \oeta Q\eta)=\exp \oeta Q_{\rF} \eta.
\end{equation}
For all $Q$ in $\symF$, $L_Q\cap W^o_\rF=\{0\}$; hence, the
symplectic reduction $t_\rF$ is smooth on $\symF\subset \lll_F$.
This is coherent with the fact that $R_{F\rightarrow \rF}(\exp
\oeta Q\eta)$ does not vanish for $Q$ in $\symF$ (cf. formula
(\ref{RFrF})).

\subsection{Intersection of the set of indeterminacy points by a holomorphic curve}

Let $U\subset \BC$ be an open subset containing 0. Let
$L:U\rightarrow \lll_F$ be analytic and such that $L(0)=L_0$ is in
the set of indeterminacy points of $\tilde t_W$, i.e. such that
$\dim(L_0\cap W^o)=n_0>0$. We suppose that $L(U)$ is not contained
in the set of indeterminacy points of $\tilde t_W$, and we may as
well suppose that $L(\lambda)$ intersects the set of indeterminacy
points at 0 only, by taking $U$ small enough. Since $\lll_F$ is
compact, $(t_W\circ L)_{|U\setminus \{0\}}$ can be analytically
continued to $U$. We choose as in the proof of proposition
\ref{p.3.RW}, ii), an orthonormal basis $\{f_1, \ldots ,f_K\}$ of
$L_0$, such that $\{f_1,\ldots ,f_{n_0}\}$ is a basis of $L_0\cap
W^o$. We can identify the tangent plane of $\lll_F$ at $L_0$ with
$\sym_K(\BC)$, thanks to the local parametrization described in
section 1.2, associated with the basis $(f_1, \ldots ,f_K)$. Let
$Z\subset \sym_K(\BC)$ be the homogeneous analytic set given by
$$
Z=\{Q\in \sym_K(\BC), \;\;\; \det((Q_{i,j})_{i,j=1}^{n_0})=0\}.
$$
\begin{lem}\label{l.3.Llambda}
If $L'(0)\in \sym_K(\BC)\setminus Z$ then
$$
\ord (\lambda\mapsto R_W\circ s(L(\lambda)), 0)=n_0,
$$
for any local section $s$ of $\pi$, in a neighborhood of $L_0$.
Moreover, $(t_W\circ L)_{|U\setminus \{0\}}$ is analytically
continued at 0 by $t_W(L_0)$.
\end{lem}
\begin{remark}
This means in particular, that if the map $\lambda\mapsto
L(\lambda)$ intersects the indeterminacy point $L_0$ in a generic
direction, then the analytic continuation is given by the point
$t_W(L_0)\in \tilde t_W(L_0)$. This points out the specific role
of the symplectic reduction $t_W(L_0)$ in the blow-up $\tilde
t_{W}(L)$.
\end{remark}
Proof: We take the notations of the proof of proposition
\ref{p.3.RW}, ii). If $L'(0)=Q_0\in\sym_K(\BC)\setminus Z$, then
in a neighborhood of 0,
$$
L(\lambda)=\vect\{f_i^{Q(\lambda)}\}_{i=1}^K,
$$
for a holomorphic function $Q(\lambda)$, with $Q(\lambda)=\lambda
Q_0+O(\lambda^2)$. From formula (\ref{f.3.RW}) we have
\begin{eqnarray*}
&& R_W(f_1^{Q(\lambda)}\wedge \cdots \wedge f_K^{Q(\lambda)})
\\
&=& C\lambda^{n_0}\det((Q_0)_{i,j=1}^{n_0}) \left(
p_{W/W^o}(f_{n_0+1})\wedge \cdots \wedge
p_{W/W^o}(f_{n_0+p+1})\right) + O(\lambda^{n_0+1}).
\end{eqnarray*}
But
$$
\pi(p_{W/W^o}(f_{n_0+1})\wedge \cdots \wedge
p_{W/W^o}(f_{n_0+p}))=t_W(L_0).
$$
Thus, $\lim_{\lambda \to 0} \tilde
t_W(L(\lambda))=t_W(L_0)$.$\diamondsuit$

\subsection{Siegel upper half-plane}
We now prove a specific property of the symplectic reduction when
the coisotropic space $W$ is the complexification of a real
subspace. Let us first introduce some definitions. The subset $S_{+,K}$
of $\sym_K(\BC)$ defined by
$$
S_{+,K}=\{Q, \;\;\; \im (Q) \hbox{ is positive definite}\},
$$
is called the Siegel upper-half plane (cf. \cite{Siegel}), and is
a homogeneous space (isomorphic to $\hbox{sp}(K,\BR)\backslash
U(K)$).

Let us remark now that for any $X$ in $V$, $\w(\overline X,X)$ is
a pure imaginary number, since $\w$ is antisymmetric. Let us
define the subset $S_{+,V}\subset \lll_V$ by
$$
S_{+,V}=\{L \in \lll_V, \;\;\; -i\w(\overline X,X)>0, \; \forall
X\in L\setminus \{0\}\}.
$$
We have then the following simple result.
\begin{propos}\label{p.3.S+}
Let $L_1$ be the complexification of a real Lagrangian subspace.
Let $v_1,\ldots ,v_K$ be a real orthonormal basis of $L_{1}$ and
$(v_1^*, \ldots, v_K^*)=J(v_1,\ldots, v_K)$ be the associated basis of
$L_1^\perp=JL_1$. For any $Q$ in $\sym_K(\BC)$ we set
$$
v_i^Q=v_i+\sum_{j=1}^K Q_{i,j} v_j^*
$$
and we  denote by $L^{(v_1,\ldots ,v_K)}_Q\in \lll_V$ the
Lagrangian subspace generated by the family $\{v_i^Q\}_{i=1}^K$.
Then we have
$$
S_{+,V} =\{L_Q^{(v_1,\ldots ,v_K)}, \;\;\; Q\in S_{+,K}\}.
$$
\end{propos}
\begin{remark}
In particular, for the canonical decomposition $V=\BC^F\oplus
(\BC^F)^*$ and the canonical basis $(e_1, \ldots ,e_K, e_1^*,
\ldots ,e_K^*)$ this gives a canonical identification of $S_{+,V}$
with $S_{+,K}$ given by
\begin{eqnarray*}
S_{+,K}&\rightarrow & S_{+,V}
\\
Q&\mapsto &L_Q,
\end{eqnarray*}
where $Q\mapsto L_Q$ is the embedding described in section 1.2
(with the notations of the previous proposition, we have
 $L_Q=L_Q^{(e_1,\ldots ,e_K)}$).
Thus, when no ambiguity is possible, we simply write $S_+$ for
$S_{+,K}\simeq S_{+,V}$.
\end{remark}
Proof: If $X$ is in $L_Q^{(v_1, \ldots ,v_K)}$ then
$$
X= \sum_{i=1}^K c_i(v_i+\sum_{j=1}^K Q_{i,j} v_j^*),
$$
for some vector $(c_1,\ldots ,c_K)\in \BC^K$. We easily get that
\begin{eqnarray}
\nonumber
\w(\overline X,X)&=&<JX,X>
\\ \label{f.3.S+}
&=& 2i \sum_{k,k'=1}^K \overline{c}_k \im(Q)_{k,k'} c_{k'}.
\end{eqnarray}
Hence, if $Q$ is in $S_{+,K}$ then $L^{(v_1,\ldots ,v_K)}_Q$ is in
$S_{+,V}$.

Conversely, we first remark that
$$
\lll_V\setminus\{L_Q^{(v_1,\ldots,v_K)},\;Q\in
\sym_K(\BC)\}=\{L\in\lll_V,\;\; L\cap L_1^\perp\neq \{0\}\}.
$$
Since $L_1$ is the complexification of a real Lagrangian subspace,
if $X$ is in $L_1^\perp=JL_1$ then so is $\overline X$. Thus
$\w(\overline X,X)=0$ for all $X$ in $L_1^\perp$, since
$L^\perp_1$ is Lagrangian. This implies that for any Lagrangian
subspace $L$ in $S_{+,V}$, $L\cap L_1^\perp=\{0\}$, and thus that
any $L$ in $S_{+,V}$ can be written $L_Q^{(v_1,\ldots ,v_K)}$ for
a certain symmetric operator $Q$ in $\sym_K(\BC)$. By formula
(\ref{f.3.S+}), we get the result. $\diamondsuit$

Let $W$ be the complexification of a real coisotropic subspace of
dimension $K+p$.
\begin{propos} \label{p.3.tWS+}
The symplectic reduction $t_W$ is analytic on $S_{+,V}$ and
$$
t_W(S_{+,V})\subset S_{+, W/W^o}.
$$
\end{propos}
Proof: Let us consider $L$ in $S_{+,V}$. As previously, we denote
by $p_{W/W^o}:W\rightarrow W/W^o$ the canonical projection. With
the identification $W/W^o\simeq W\cap JW$, we can write any point
$X$ in $L\cap W$ as $X=p_{W/W^o}(X)+X'$ with $X'\in W^o$. Using the fact
that $W$ is the complexification of a real subspace we have
$p(\overline X)=\overline{p(X)}$ and $\overline{W^o}=W^o$. Thus,
for any vector $X$ in $L\cap W$
\begin{eqnarray*}
\w(\overline X,X)&=& \w(p(\overline{ X}),p(X))+
\w(p(\overline{X}),{X'})+\w(\overline{X}',p(X))+\w(\overline X',X)
\\
&=& \w(\overline{p(X)},p(X)),
\end{eqnarray*}
since $W^o\perpw W$. This implies that $-i\w(\overline
{p(X)},p(X))>0$ for all $X$ in $L\cap W\setminus\{0\}$. We deduce,
firstly, that $p(X)\neq 0$ for all $X\neq 0$ in $L\cap W$, thus
that $L\cap W^o=\{0\}$, and that $t_W$ is continuous on $S_{+,V}$.
Secondly, we deduce that $t_W(L)$ is in $S_{+,W/W^o}$.
$\diamondsuit$

\subsection{A special class of holomorphic curves}

When $W$ is real, there is a natural class of applications
$\lambda\mapsto L(\lambda)$ which satisfies the hypotheses of
lemma \ref{l.3.Llambda} at any point.
\begin{lem}\label{l.3.LS+}
Let us suppose that $L:\BC\rightarrow \lll_F$ is holomorphic on
$\BC$ and such that
$$
L(\BR)\subset \lll_{F,\BR}, \;\hbox{ and, }\; L(\{\lambda, \im
\lambda
>0\})\subset S_{+,V},
$$
then $\lambda\mapsto L(\lambda)$ satisfies the hypotheses of lemma
\ref{l.3.Llambda} at any point of $\BC$, i.e. $\lambda\mapsto
t_W(L(\lambda))$ is holomorphic and
$$
\ord(\lambda \mapsto R_W\circ s(L(\lambda)), \lambda_0)=\dim (
L(\lambda_0)\cap W^o),
$$
if $s$ is a local holomorphic section of $\pi$ on a neighborhood
$U\subset \lll_F$ of $L(\lambda_0)$.
\end{lem}
Proof: Clearly, $L(\lambda)$ may intersect the set of
indeterminacy points of $t_W$ at real points only. Hence,
$\lambda\mapsto L(\lambda)$  satisfies the hypotheses of lemma
\ref{l.3.Llambda} on $\BC\setminus \BR$. Let $\lambda_0$ be real,
and $(v_1, \ldots ,v_K)$ be a real orthonormal basis of
$L(\lambda_0)$, and $(v_1^*, \ldots ,v^*_K)={J(v_1, \ldots
,v_K)}$. In a neighborhood of $\lambda_0$, we have
$L(\lambda)=L_{Q(\lambda)}^{(v_1, \ldots ,v_K)}$, with the
notations of proposition \ref{p.3.S+}, and with a holomorphic
$Q(\lambda)$. Since $L(\BR)\subset \lll_{F,\BR}$, $Q'(\lambda_0)$
is real. It is also positive definite. Indeed, let $C=(c_i)$ be in
$\BR^K\setminus \{0\}$ and consider $X_\lambda$ in $V$ given by
$$
X_\lambda=\sum_{i=1}^K c_i(v_i+\sum_{j=1}^K (Q(\lambda))_{i,j} v_j^*).
$$
Then we have, cf. formula (\ref{f.3.S+}),
$$
-i\w(\overline{X_\lambda}, X_\lambda)=2 \im(<C, Q(\lambda) C>).
$$
On the other hand, we have locally the following approximation
$$
<C, Q(\lambda) C>\;= {(\lambda-\lambda_0)^n\over n!}
<C,Q^{(n)}(\lambda_0)C> +\; O((\lambda-\lambda_0)^{n+1}),
$$
where $Q^{(n)}(\lambda_0)$ is the $n$-th derivative of
$Q(\lambda)$ and $n$ the smallest integer such that
$<C,Q^{(n)}(\lambda_0) C> \;\neq 0$ (this $n$ exists, since
otherwise $<X, Q(\lambda) X>\;=0$ for all $\lambda$). For all
$\lambda$ such that $\im \lambda >0$, we must have $\im (<C,
Q(\lambda) C>)>0$ since $X_\lambda$ is in $L(\lambda)$, and
$L(\lambda)$ in $S_{+,V}$. This is possible only if $n=1$ and $<C,
Q'(\lambda_0) C>>0$. Hence, $Q'(\lambda_0)$ is positive definite.
This immediately implies that $\lambda\mapsto L(\lambda)$
satisfies the hypothesis of lemma \ref{l.3.Llambda} for all
$\lambda_0$ in $\BC$.$\diamondsuit$

\section{Application to the renormalization map of finite
self-similar structures}
\subsection{Finite self-similar structures}
Here, we introduce the notion of finite self-similar structures.
These structures appear in relation with finitely ramified
self-similar sets (also called p.c.f self-similar sets), cf.
section 5.1. A generalized version of these structures also seems
to appear in relation with some automatic groups, cf. section 5.2
and examples, section 7.

Let $F=\{1,\ldots ,K\}$ be a finite set and $N$ an integer, $N\ge
2$. We set
$$
\tilde F_\un =\unN\times F,
$$
and $F_{\un, i}=\{i\}\times F\subset \tilde F_\un$.

We suppose given an equivalence relation $\rrr$ on $\tilde F_\un$
and we set
$$
F_\un =\tilde F_\un/\rrr.
$$
We denote by $\pi:\tilde F_\un\rightarrow F_\un$ the canonical
projection. Finally, we suppose that a subset $\pF_\un$ is
specified in $F_\un$, together with a bijective map between $F$
and $\pF_\un$, which gives a canonical identification between $F$
and $\pF_\un$ (cf. the example of the Sierpinski gasket in the
introduction and in section 7).
 \ali
We call finite self-similar structure, the triplet $(F,N,\rrr)$
together with the identification of a subset $\partial
F_\un\subset F_\un$ with $F$.

From this finite structure, we can construct a sequence of sets
$F_\nn$, with an identification of a subset $\pF_\nn\subset F_\nn$
with $F$, as follows. Suppose that the sequence $(F_\nn, \pF_\nn)$
is constructed up to level $n$. We consider the set $\unN\times
F_\nn$; the subset $\unN\times \pF_\nn$ can be identified with
$\tilde F_\un$; then inside $\unN\times F_\nn$, we glue together
the points of $\unN\times \pF_\nn$ according to the relation
$\rrr$. This gives a set $F_{<n+1>}$ which contains a copy of
$F_\un$: thus, we define $\pF_{<n+1>}$ as the boundary set
$\pF_\un$, when $F_\un$ is considered as the subset $\unN\times
\pF_\nn/\rrr$. Remark that $F_\nn$ can also be considered as the
quotient of
$$
\tilde F_\nn=\unN^n\times F
$$
by an equivalence relation that we denote $\rrr_\nn$ (but which
we do not describe explicitly here). We denote by $\tilde
F_{\nn,i_1, \ldots ,i_n}$ (resp. $F_{\nn,i_1,\ldots ,i_n}$) the
subset of $\tilde F_\nn$ (resp. of $F_\nn$) of the type
$\{(i_1,\ldots ,i_n)\}\times F$ (resp. $\{(i_1,\ldots
,i_n)\}\times F/\rrr_\nn$).

\subsection{Self-similar Schr\"odinger operators}
Let $\rho=((\rho_{i,j})_{i\neq j}, (\rho_i)_{i\in F})$, be a
dissipative electrical network on $F$, as defined in section 1.1.
We define an electrical network $\tilde \rho_\nn$ on $\tilde
F_\nn=\unN\times F$ by making a copy of $\rho$ on each $\tilde
F_{\nn, i_1, \ldots ,i_n}$.
Otherwise stated, it means that
\begin{equation}
\label{equa-rho}
\tilde\rho_{|F_{\nn, i_1, \ldots ,i_n}}=\rho, \;\;\;\forall
i_1,\ldots ,i_n,
\end{equation}
and that the conductances between two different
subsets $F_{\nn, i_1, \ldots ,i_n}$ and $F_{\nn, i'_1, \ldots ,i'_n}$
are null.
Then we define $\rho_\nn$ as the
electrical network on $F_\nn$ obtained from $\tilde \rho_\nn$ by
the gluing map described in section 2.2 (considering that
$F_\nn=\tilde F_\nn/\rrr_\nn$).

Similarly, if $b$ is a positive measure on $F$, it induces a
positive measure $\tilde b_\nn$ on $\tilde F_\nn$ equal to $b$ on
each subset $\tilde F_{\nn,i_1,\ldots ,i_n}$, and a positive
measure $b_\nn$ on $F_\nn$, image of $\tilde b_\nn$ by the
canonical projection $\tilde F_\nn\rightarrow F_\nn$.
 \ali

Let $H_\nn=H_{\nn, \rho,b}$ be the Schr\"odinger operator defined
from $(\rho_\nn,b_\nn)$ by
$$
<Q_{\rho_\nn} f,h>=-\int H_\nn (f)(x) h(x) db_\nn(x), \;\;\;
\forall f,h\in \BR^{F_\nn}.
$$
N.B.: $<,>$ is the usual scalar product on $\BR^{F_\nn}$.

We denote by $\nu_\nn^+$ the counting measure of the eigenvalues
of $H_\nn$. We denote by $\nu_\nn^-$ the counting measure of the
Dirichlet eigenvalues of $H_\nn$, i.e. of the eigenvalues of the
restriction of $H_\nn$ to
$$
\ddd_\nn^-=\{f\in\BR^{F_\nn}, \;\; f_{|\pF_\nn}=0\}.
$$
Remark that $f\in \BR^{F_\nn}$ is an eigenfunction of $H_\nn$ with
eigenvalue $\lambda$ if and only if
\begin{eqnarray}
\label{f.new.1} (Q_{\rho_\nn} +\lambda I_{b_\nn})f=0,
\end{eqnarray}
where $I_{b_\nn}$ is the diagonal operator with diagonal terms
$(I_{b_\nn})_{x,x}=b_\nn(x)$ for all $x$ in $F_\nn$. Similarly,
$f$ is a Dirichlet eigenfunction with eigenvalue $\lambda$ if
and only if
\begin{eqnarray}
& \label{f.new.2} \left( (Q_{\rho_\nn} +\lambda
I_{b_\nn})f\right)_{|F_\nn\setminus \pF_\nn}=0,
\\
& \label{f.new.3} f_{|\pF_\nn}=0.
\end{eqnarray}
We denote by $\mu$ the following limit
$$
\mu=\lim_{n\to \infty} {1\over N^n} \nu_\nn^\pm,
$$
when it exists and does not depend on the boundary condition
$\pm$. The measure $\mu$ is called the density of states (the
existence of this limit was first proved in \cite{Fukushima2},
\cite{KigamiL}).

Let us now define the so-called Neumann-Dirichlet eigenvalues. A
function $f:{F_\nn}\rightarrow \BR$ is a Neumann-Dirichlet (N-D
for short) eigenfunction of $H_\nn$ with eigenvalue $\lambda$ if
$H_\nn f=\lambda f$ and $f_{|\pF_\nn}=0$. This means that $f$ is
both a Neumann and Dirichlet eigenfunction. Obviously, $f$ is a
N-D eigenfunction of $H_\nn$ with eigenvalue $\lambda$ if and only
if $f$ satisfies (\ref{f.new.1}) and (\ref{f.new.3}).
 We denote by $\nu_\nn^\ND$ the
counting measure (with multiplicity) of the N-D eigenvalues. It is
clear that if $f$ is a N-D eigenfunction on $F_\nn$, then we can
make $N$ independent copies of $f$ on each subcell of $F_{<n+1>}$
(cf. \cite{Sabot5}). Thus,
$$
\nu_{<n+1>}^\ND\ge N\nu_\nn^\ND,
$$
and the limit
$$
\mu^\ND=\lim_{n\to\infty}{1\over N^n} \nu_\nn^\ND,
$$
exists and is called the density of Neumann-Dirichlet states.
 These two measures play an important role in the
understanding of the spectral properties of some infinite
self-similar lattices, cf. \cite{Sabot6}, and section 5.

\subsection{The renormalization map}
In \cite{Sabot5}, we have introduced a renormalization map defined
on the Lagrangian Grassmannian $\lll_F$. We recall its definition
and give its expression in terms of a symplectic reduction. In
particular, we give an explicit expression of the zeros of the
associated map on the Grassmann algebra, and this can be useful
for applications.

Firstly, this renormalization map can be defined on $\symF(\BC)$
as follows. Let $Q$ be in $\symF(\BC)$. We denote by $\tilde
Q_\un$ the block-diagonal symmetric operator on $\BC^{\tilde
F_\un}$ defined by
$$
\tilde Q_\un=Q\oplus\cdots \oplus Q,
$$
on the decomposition $\BC^{\tilde F_\un}=\BC^{\tilde
F_{\un,1}}\oplus \cdots \oplus \BC^{\tilde F_{\un,N}}$. We denote
by $Q_\un$ the symmetric operator defined on $\BC^{F_\un}$ by the
gluing operation described in section 1.2. Then we take the trace
operator $(Q_{\un})_{\pF_\un}$, which is a symmetric operator on
$\BC^{\pF_\un}\sim\BC^F$.

We denote  by $T: \sym_F(\BC)\rightarrow \sym_F(\BC)$ the map
given by
$$ T(Q)=(Q_\un)_{\pF_\un}.
$$
The coefficients of $T(Q)$ are rational functions of the
coefficients of $Q$ and the poles are included in the set
$\det(Q_{|F_\un\setminus \pF_\un})=0$. It is clear that the
iterate $T^n$ has the following expression: let  $\tilde Q_\nn$ be
the block-diagonal symmetric operator on $\BC^{\tilde F_\nn}$
defined by
$$
\tilde Q_\nn=Q\oplus\cdots \oplus Q,
$$
on the decomposition $\BC^{\tilde F_\nn}=\oplus_{i_1, \ldots ,i_n}
\BC^{\tilde F_{\nn,i_1, \ldots ,i_n }}$ (cf. the notations of
section 4.1), and $Q_\nn$ the element of $\sym_{F_\nn}$ obtained
by gluing from $\tilde Q_\nn$ (considering that $F_\nn$ is a
quotient of $\tilde F_\nn$). Then we have
$T^n(Q)=(Q_\nn)_{\pF_\nn}$.

The map $T$ is the composition of three operations: the map
$Q\mapsto \tilde Q_\un$, the gluing map $\tilde Q_\un\mapsto
Q_\un$ and the trace map $Q_\un\mapsto (Q_\un)_{\pF_\un}$. The
last two operations correspond to symplectic reductions on the
Lagrangian compactification. Since the composition of two
simplectic reductions is a simplectic reduction we see that the
extension of this map to the Lagrangian compactification must be
represented by the composition of the ``copies" map with a
symplectic reduction.

Let us describe precisely this map on the Lagrangian
compactification. We set $V_{\tilde F_\un}=\BC_{\tilde
F_\un}\oplus (\BC_{\tilde F_\un})^*$ and we have the decomposition
$$V_{\tilde F_\un}=V_{\tilde F_{\un,1}}\oplus \cdots \oplus V_{\tilde
F_{\un,N}},
$$
with obvious notations. For any $L$ in $\lll_F$, we denote by
$\tilde L_\un$ the Lagrangian subspace of $V_{\tilde F_\un}$ equal
to $\tilde L_\un=L\oplus \cdots \oplus L$. It is clear that the
map $L\mapsto \tilde L_\un$ extends  the map $Q\mapsto \tilde
Q_\un$ to the Lagrangian compactifications $\lll_F\rightarrow
\BL_{\tilde F_\un}$. Considering section 2.1 and 2.2, we see that
the map $g:\lll_F\rightarrow \lll_{\pF_\un}\simeq \lll_F$ defined
by
$$
g(L)=\tilde t_{F_\un\rightarrow \pF_\un}\circ \tilde t_{\tilde
F_\un\rightarrow F_\un}\circ (L\mapsto \tilde L_\un),
$$
extends the map $T$ to the Lagrangian compactification.
\begin{remark}
Formally, the map $g$ is defined as the rational map obtained as a
composition of rational maps, as defined at the beginning of
section 3.1. Remark that this composition is well-defined since
the Siegel upper half-space $S_{+,V}$, which is an open subset of
$\BL_F$, is preserved by all these maps, and does not intersect
the indeterminacy points of the symplectic reductions.
\end{remark}
 From
section 1.3, we know that the composition $t_{F_\un\rightarrow
\pF_\un}\circ t_{\tilde F_\un\rightarrow F_\un}$ can be expressed
directly as a symplectic reduction. Let us denote by
$s:\BC^{F_\un}\rightarrow \BC^{\tilde F_\un}$ the canonical
injection given by $s(f)=f\circ\pi$, and by $s^*:(\BC^{\tilde
F_\un })^*\rightarrow (\BC^{F_\un})^*$ the dual linear map (i.e.
the map given by $ s^*(e^*_x)= e^*_{\pi(x)}$).  We consider the
subspace $W_\un\subset V_{\tilde F_\un}$ defined by
\begin{eqnarray*}
W&=&  \im(s)\oplus (s^*)^{-1}((\BC^{\pF_\un})^*)
\\
&\simeq& \BC^{F_\un}\oplus \left( (\BC^{\pF_\un})^*\oplus
\ker(s^*)\right).
\end{eqnarray*}
It is clear that the $\w$-orthogonal subspace $W_\un^o$ is equal
to
$$
W_\un^o=s(\BC^{F_\un\setminus \pF_\un})\oplus \ker(s^*)\simeq
\BC^{F_\un\setminus \pF_\un}\oplus \ker(s^*).
$$
Hence, $W_\un$ is coisotropic and $W_\un/W_\un^o$ is isomorphic to
the symplectic structure $V_{\pF_\un}\sim V_F$ . From formula
(\ref{composition}), we know that $t_{F_\un\rightarrow
\pF_\un}\circ t_{\tilde F_\un\rightarrow F_\un}=t_{W_\un}$, and thus that
$\tilde t_{F_\un\rightarrow
\pF_\un}\circ \tilde t_{\tilde F_\un\rightarrow F_\un}=\tilde t_{W_\un}$.
Hence, $g$ has the following expression
\begin{eqnarray*}
g: \lll_F&\rightarrow & \lll_F
\\
L&\mapsto & \tilde t_W\circ (L\mapsto \tilde L_\un).
\end{eqnarray*}
The iterates $g^n$ can be described in a similar fashion. we have
the decomposition
$$ V_{\tilde F_\nn}=\oplus_{i_1,\ldots ,i_n}
V_{\tilde F_{\nn,i_1, \ldots ,i_n}}.
$$
 For $L$ in $\lll_F$ we define $\tilde L_\nn$ as the element of
$V_{\tilde F_\nn}$ given by
$$
\tilde L_\nn =L\oplus \cdots \oplus L.
$$
Then it is easy to see that
$$
g^n(L)=\tilde t_{F_\nn\rightarrow \pF_\nn}\circ \tilde t_{\tilde
F_\nn\rightarrow F_\nn}\circ (L\mapsto \tilde L_\nn).
$$
As previously, we denote by $s_\nn:\BC^{F_\nn}\rightarrow
\BC^{\tilde F_\nn}$ the natural linear injection, and by $s_\nn^*$
the dual linear operator. We set
\begin{eqnarray*}
W_\nn&=& \im(s_\nn)\oplus (s_\nn^*)^{-1}((\BC^{\pF_\nn})^*)
\\
&\simeq& \BC^{F_\nn}\oplus \left( (\BC^{\pF_\nn})^*\oplus
\ker(s_\nn^*)\right).
\end{eqnarray*}
The subspace $W_\nn$ is coisotropic and we have
$$
W_\nn^o= s(\BC^{F_\nn\setminus \pF_\nn})\oplus \ker(s_\nn^*).
$$
As previously, we have
$$
g^n(L)=\tilde t_{W_\nn}\circ (L\mapsto \tilde L_\nn).
$$

 \ali
{\it The corresponding map on the Grassmann algebra}
 \ali
We recall that $\aaa_F$ is the Grassmann algebra associated with
the set $F$, described in section 1. The smooth manifold $\lll_F$
is a projective variety and is embedded in $\ppp(\aaa_F)$, cf.
section 1.4. The map $g$ can be lifted to a homogeneous polynomial
map on $\pi^{-1}(\lll_F)\subset \aaa_F$, that we describe now. If
$X$ is in $\aaa_F$ we denote by $\tilde X_\un$ the element of
$\aaa_{\tilde F_\un}$ defined by
$$
\tilde X_\un=X_{\un,1}\cdots X_{\un,N},
$$
where $X_{\un,i}$ is the element of $\aaa_{F_{\un,i}}$
corresponding to $X$ in $\aaa_F$ (indeed, $\aaa_{F_{\un,i}}\simeq
\aaa_F$). Then we set
\begin{eqnarray*}
R:\aaa_F&\rightarrow & \aaa_{\pF_\un}\sim\aaa_F
\\
X&\mapsto & R_{F_\un\rightarrow \pF_\un}\circ R_{\tilde
F_\un\rightarrow F_\un}(\tilde X_\un),
\end{eqnarray*}
where $R_{F_\un\rightarrow \pF_\un}$ and  $R_{\tilde
F_\un\rightarrow F_\un}$ are the maps defined in section 2.1 and
2.2 respectively. It is clear that the map $X\mapsto \tilde X_\un$
is a homogeneous polynomial map of degree $N$ in the coefficients
of $X$ and that it lifts the map $L\mapsto \tilde L_\un$ to
$\pi^{-1}(\lll_F)$, i.e. that $\pi(\tilde X_\un)=\tilde L_\un$. It
is clear by section 2 and 3, that $R_{F_\un\rightarrow
\pF_\un}\circ R_{\tilde F_\un\rightarrow F_\un}$ corresponds on
$\aaa_F$ to the map $R_{W_\un}$ defined in section 3, and thus
that it lifts the symplectic reduction $\tilde t_{W_\un}$ to
$\pi^{-1}(\lll_{\tilde F_\un})$. Hence, we have shown the
commutation of the following diagram
$$
\begin{CD}
{\pi^{-1}(\lll_F)} @>\mathrm{R}>> {\pi^{-1}(\lll_F)}\\
@VV{\mathrm{\pi}}V @VV{\mathrm{\pi}}V\\
{\lll_F}@>\mathrm{g}>> {\lll_F}
\end{CD}
$$
on the set where these maps are well-defined. Let us finally
remark that the map $R$ is a homogeneous polynomial map of degree
$N$, since $R_{F_\un\rightarrow \pF_\un}$ and $R_{\tilde
F_\un\rightarrow F_\un}$ are linear. Similarly, we have
$R^n(X)=R_{F_\nn\rightarrow \pF_\nn}\circ R_{\tilde
F_\nn\rightarrow F_\nn}(\tilde X_\nn)$, where $\tilde X_\nn$ is
the product $\tilde X_\nn= \prod_{i_1, \ldots ,i_n}
X_{\nn,i_1,\ldots ,i_n}$ (with obvious notations). This formula is
simple (and proved in detail in \cite{Sabot5}, proposition 3.1,
ii)).

\subsection{Group of symmetries}
Most of the classical examples have a natural group of symmetries,
that we denote $G$. In these cases it is natural to restrict our
analysis to $G$-invariant objects, i.e. $G$-invariant electrical
networks, $G$-invariant measures and to consider the restriction
of the map $g$ to $G$-invariant Lagrangian subspaces. The
properties of the map $g$ can be quite different from those of its
restriction.
\begin{remark}
This section is not necessary to understand the rest of the text
(with the exception of section 7), and we advise a reader not
familiar with the subject to skip it and forget any
 reference
about the group $G$, upon a first reading.
\end{remark}

This invariance by a group of symmetries
can be formalized as follows. We suppose given a finite group
$G$ together with an action of $G$ on $\unN$ and $F$. We suppose
that the equivalence relation $\rrr$ is compatible with the action
of $G$ on $\unN\times F$, i.e. that
$$
(g\cdot i, g\cdot x)\rrr (g\cdot j, g\cdot y) \;\; \hbox{iff} \;\;
(i,x)\rrr(i,y).
$$
This induces an action of $G$ on the quotient set $F_\un$, and we
suppose that the subset $\pF_\un$ is left invariant by the action
of $G$ and that the identification between $F$ and $\pF_\un$
commutes with the action of $G$ on $F$ and $\pF_\un$. This implies
that the action of $G$ can be defined on all $F_\nn$, by the
action of $G$ on $\unN^n\times F$ (cf. the example of the
Sierpinski gasket).

We denote by $\sym^G_F(\BC)$, the subspace of $G$-invariant
elements of $\sym_F(\BC)$, i.e. the space of symmetric matrices
$Q$ which satisfies
$$
Q_{g\cdot i, g\cdot j}=Q_{i,j}, \;\;\;\forall g\in G.
$$
 The space $\sym^G_F(\BC)$ is embedded in $\BL_F$, since
 $\sym_F(\BC)$ is embedded in $\BL_F$. We denote by $\BL_F^G$, the
 closure in $\BL_F$ of $\sym_F^G(\BC)$. As shown in \cite{Sabot5},
 appendix E, $\BL^G_F$ is included in the subset of $G$-invariant
Lagrangian subspaces of $V$, and is a smooth projective variety,
which can be locally parametrized by $\sym_F^G(\BC)$.
 Obviously, the map $ T$ sends $\sym^G_F(\BC)$ to itself, since
 the whole structure is $G$-invariant, and thus,
$g:\lll_F\rightarrow \lll_F$ leaves the subvariety $\lll_F^G$
invariant. If we are only interested in $G$-invariant
Schr\"odinger operators, then it is much better to consider only
the map on the invariant subvariety $\lll_F^G$. In particular,
some values like the asymptotic degree of the map might be very
different on $\lll_F$ and on $\lll_F^G$, and the significant value
comes from the map on $\lll_F^G$. In the following, we will always
consider the restricted map $g: \BL_F^G\rightarrow \BL_F^G$.
\begin{remark}
Remark that $\lll_F^G$ is not contained in the set of
indeterminacy points of $g$ since $\lll_F^G\cap S_{+,V}\neq
\emptyset$. Hence, the restriction of $g$ to $\BL_F^G$ is well-defined.
\end{remark}

In \cite{Sabot5}, appendix E, we have described explicitly the
structure of $\BL^G_F$. This structure depends on the type and the
multiplicities of the real representation of $\BR^F$. Indeed, the
space $\BR^F$ is the sum of $r$ distinct irreducible real
representations $W_0, \cdots ,W_r$, with multiplicities $n_0,
\cdots ,n_r$. In \cite{Sabot5}, we have proved that $\BL^G_F$ is
isomorphic to the product
$$
{\mathcal L}_0\times \cdots \times {\mathcal L}_{r},
$$
where the ${\mathcal L}_i$ are Grassmannians of one of the
following three types: Lagrangian Grassmannian (as defined in
section 1), Grassmannian of $k$-dimensional subspaces in
$\BC^{2k}$, or orthogonal Grassmannian (cf. \cite{Sabot5}). The
type of ${\mathcal L}_i$ depends on the type of the representation
$W_i$, and the dimension depends on the multiplicity $n_i$. The
main point is that, independently of the type and the multiplicity
of ${W}_i$, we have
$$
\dim( H^1({\mathcal L}_i))=0, \;\;\; \dim(H^2({\mathcal L}_i))=1,
$$
where $H^k({\mathcal L}_i)$ is the $k$-th cohomology group of ${\mathcal
L} _i$.

\subsection{Properties of the map $g$ and $R$}
The following properties are easy consequences of section 3.
\begin{propos}\label{p.4.1}

i) The set of indeterminacy points of $g$ does not intersect
$S_{+,V}$ and
$$
g(S_{+,V})=S_{+,V}.
$$
 \ali
ii) For any $L_0$ in $\lll_{F,\BR}^G$
$$
\ord( R^n\circ s, L_0)=\dim((\widetilde{L_0})_\nn\cap W^o_\nn),
$$
where $s$ is a holomorphic section of the projection $\pi$ on a
neighborhood $U\subset \lll_F^G$ of $L_0$.
 \ali
iii) Let $L:\BC\rightarrow \lll_F^G$ be  holomorphic and such that
\begin{eqnarray} \label{f.4.ord}
L(\BR)\subset \lll_{F,\BR}^G, \;\hbox{ and }\; L(\{\lambda, \im
\lambda >0\})\subset S_+.
\end{eqnarray}
At any $\lambda_0$ in $\BC$, we have
\begin{eqnarray}\label{f.4.ordlambda}
\ord( \lambda\mapsto R^n\circ s (L(\lambda)), \lambda_0)= \dim(
(\widetilde{L(\lambda_0)})_\nn\cap W_\nn^0).
\end{eqnarray}
Moreover, the map $\lambda \mapsto
t_{W_\nn}((\widetilde{L(\lambda)})_\nn)$ gives a holomorphic
extension of $\lambda\mapsto g^n(L(\lambda))$ at any point where
$L(\lambda)$ intersects the set of indeterminacy points of $g^n$.
\end{propos}
\begin{remark}
By the theorem of resolution of singularities, since $\BL_F$ is
compact, we know that $g^n(L)=\tilde
t_{W_\nn}((\widetilde{L(\lambda)})_\nn)$ has a holomorphic
extension to $\BC$. The point iii) says that at any point of
indeterminacy of $g^n$ this holomorphic extension is given by the
symplectic reduction $t_{W_\nn}((\widetilde{L(\lambda)})_\nn)$
(recall that the symplectic reduction is defined everywhere).
\end{remark}
\begin{remark}
 A priori, the order of vanishing on the left hand side of
(\ref{f.4.ord}) and (\ref{f.4.ordlambda}) is bigger than the
right-hand side, from proposition \ref{p.3.RW}. For real points
and for the restriction to some specific curves, we see that there
is actually equality. The type of holomorphic curves that appears
in ii) is exactly the type of curve that we will encounter later
on.
\end{remark}
Proof: It is clear that if $L$ is in $S_{+,V}$, then $\tilde
L_\un$ is in $S_{+,\tilde V_\un}$. This implies that i) and iii)
are direct consequences of proposition \ref{p.3.tWS+} and lemma
\ref{l.3.LS+}.
 \ali
ii) Let $(v_1, \ldots ,v_K)$ be a real orthonormal basis of $L_0$.
The map $\lambda \mapsto L(\lambda)=L_{\lambda \Id}^{(v_1, \ldots
,v_K)}$ satisfies hypothesis of iii). Thus, the left hand side of
(\ref{f.4.ord}) is smaller than the right-hand side. It is also
bigger or equal by proposition \ref{p.3.RW}.$\diamondsuit$
\begin{remark}
It is trivial from the definition, that $T$ is 1-homogeneous, i.e.
that $ T(\alpha Q)= \alpha T(Q)$, for all $\alpha\in\BC$. This
1-homogeneity of $T$ induces the following invariance property of
$g$
\begin{eqnarray}\label{f.4.1homogeneity}
\tau_\alpha\circ g=g\circ \tau_\alpha,
\end{eqnarray}
for any non-null complex number $\alpha$ (N.B.: $\tau_\alpha$ is
defined in 2.3). This commutation property is non longer true for
some natural generalizations of these models, cf. section 5.2.2.
\end{remark}

\subsection{Counting measures, density of states and Green
current}
 The iterates of the map $g$ are used in \cite{Sabot5} to
describe the counting measures $\nu_\nn^\pm$ of the self-similar
Schr\"odinger operator $H_\nn$ on $F_\nn$, and the counting
measures $\nu_\nn^\ND$ of the Neumann-Dirichlet eigenvalues. We
don't want to go to much into the details here; we just want to
explain the ideas behind the main results of \cite{Sabot5}, and we
try to insist on the new perspectives given by the interpretation
in terms of symplectic reduction.

We will be interested in the holomorphic curve
\begin{eqnarray*}
\BC&\rightarrow & \lll_F
\\
\lambda &\mapsto & L_{Q_\rho+\lambda I_b}.
\end{eqnarray*}
Let us first remark that this map satisfies the hypotheses of
proposition (\ref{p.4.1}), iii), since
$\im (Q_\rho+\lambda I_b)=(\im \lambda) I_b$.
We also introduce the map
\begin{eqnarray*}
\phi:\BC&\rightarrow & \hbox{$\aaa_F$}
\\
\lambda &\mapsto & \exp \oeta (Q_\rho+\lambda I_b)\eta ,
\end{eqnarray*}
which lifts $ L_{Q_\rho+\lambda I_b}$ to $\aaa_F$, i.e.
$$
\pi ( \phi(\lambda))= L_{Q_\rho+\lambda I_b}.
$$
Let us now explain why the spectrum of $H_\nn$ is related to the
iterates $g^n( L_{Q_\rho+\lambda I_b})$ and $R^n(\phi(\lambda))$.
We first describe the Lagrangian subspace $g^n( L_{Q_\rho+\lambda
I_b})$. We already
 know that
$t_{\tilde F_\nn\rightarrow F_\nn} ((\widetilde{ L_{Q_\rho+\lambda
I_b}})_\nn))= L_{Q_{\rho_\nn+\lambda I_{b_\nn}}}$.
 For $\lambda$ in $\BC$, we consider
the following equation on $F_\nn$
\begin{equation}
\label{equa-Hn} \left( (Q_{\rho_\nn}+\lambda
I_{b_\nn})f\right)_{|F_\nn\setminus \pF_\nn}=0.
\end{equation}
For $f$ solution of the previous equation, we define
$I_f^{\rho_\nn,\lambda}$ as the element of $(\BC^{F_\nn})^*$ given
by
$$
I_f^{\rho_\nn,\lambda}(h)=\; <(Q_{\rho_\nn}+\lambda I_{b_\nn})f,
h>,\;\; \forall h \in \BR^{F_\nn}.
$$
Clearly, $I_f^{\rho_\nn,\lambda}$ is supported by $\pF_\nn$, and
thus, in $(\BC^{\pF_\nn})^*$. From formula (\ref{LQ-pF}), it is
clear that
$$
g^n(L_{Q_\rho+\lambda I_b})=\{f_{|\pF_\nn}+
I_f^{\rho_\nn,\lambda}, \;\;\hbox{ f solution of
$(\ref{equa-Hn})$}\}.
$$
Hence, we see that $g^n(L_{Q_\rho+\lambda I_b})$ contains information about
the boundary values of the space of solutions of (\ref{equa-Hn}).
In particular, we see that the number of Neumann-only eigenfunctions
(i.e. Neumann eigenfunctions which are not N-D eigenfunctions),
with eigenvalue $\lambda_0$ is given by
$$ \nu_\nn^+(\{\lambda_0\})-\nu_\nn^\ND(\{\lambda_0\})=
\dim\left( g^n(L_{Q_\rho+\lambda_0 I_b})\cap (\BC^F\oplus 0)\right).
$$
Similarly, we have
$$ \nu_\nn^-(\{\lambda_0\})-\nu_\nn^\ND(\{\lambda_0\})=
\dim\left( g^n(L_{Q_\rho+\lambda_0 I_b})\cap (0\oplus (\BC^F)^*)\right),
$$
for the Dirichlet spectrum. On the other hand, the number of N-D
eigenfunctions with eigenvalue $\lambda$ is obtained as the order
of vanishing of $R^n\circ\phi$ at $\lambda$. Indeed, from
proposition \ref{p.4.1}, we know that the order of vanishing of
$R^n(\phi(\lambda)$ is equal to the dimension of
$$
\widetilde{(L_{Q_\rho+\lambda I_b})}_\nn\cap W^o_{\nn}.
$$
But, considering that
$$
\widetilde{(L_{Q_\rho+\lambda I_b})}_\nn\cap W^o_{\tilde F_\nn/\rrr_\nn}=\{0\}
$$
since $t_{\tilde F_\nn\rightarrow F_\nn}$ is smooth on
$\sym_{\tilde F_\nn}$, we know that
$$
\widetilde{(L_{Q_\rho+\lambda I_b})}_\nn\cap W^o_{\nn}
\simeq
{L_{Q_{\rho_\nn}+\lambda I_{b_\nn}}}\cap W^o_{\pF_\nn}
$$
(indeed $t_{\tilde F_\nn\rightarrow
F_\nn}(\widetilde{(L_{Q_\rho+\lambda I_b})}_\nn)=
L_{Q_{\rho_\nn}+\lambda I_{b_\nn}}$). We have
$$
{L_{Q_{\rho_\nn}+\lambda I_{b_\nn}}}\cap W^o_{\pF_\nn}=
\ker^\ND(Q_{\rho_\nn}+\lambda I_{b_\nn})\oplus 0,
$$
where
$$
\ker^\ND(Q_{\rho_\nn}+\lambda I_{b_\nn}) =\{ f\in \BR^{F_\nn},\;\;
(Q_{\rho_\nn}+\lambda I_{b_\nn})f=0 \hbox{ and } f_{|\pF_\nn}=0\}
$$
is the subspace generated by the N-D eigenfunctions of $H_\nn$
with eigenvalue $\lambda$. Hence, we have
$$
\ord (\lambda\mapsto R^n(\phi(\lambda)), \lambda_0)= \nu_\nn^\ND
(\{\lambda_0\}).
$$
Hence, we see that the maps $g^n$ and $R^n$ contain a lot of
information about the spectrum of $H_\nn$. Remark that the
relation between $R^n$ and the spectrum of $H_\nn$ can also be seen
from the following simple formulae:
\begin{eqnarray*}
&&\det(Q_{\rho_\nn}+\lambda I_{b_\nn})=\;<R^n\circ \phi(\lambda),
X^+>,
\\
&&\det((Q_{\rho_\nn}+\lambda I_{b_\nn})_{|F_\nn\setminus
\pF_\nn})=\;<R^n\circ \phi(\lambda),X^->,
\end{eqnarray*}
where
$$ X^+=\Pi_{x\in F} \oeta_x \eta_x, \;\;\; X^-=1.
$$
(These formulas are direct consequences of the definition of
$R^n$). Hence, we see that $\nu^+(\{\lambda_0\})$ (resp.
$\nu^-(\{\lambda_0\})$) corresponds to the order of vanishing of
$\lambda \mapsto \hbox{$<R^n\circ \phi(\lambda), X^+>$}$ (resp.
$\lambda \mapsto <R^n\circ \phi(\lambda), X^->$) at $\lambda_0$.
In particular we have the following expression of $\nu_\nn^\pm$
\begin{eqnarray}
\label{formule-nun}
\nu_\nn^\pm
= {1\over 2\pi} \Delta \ln \vert <R^n\circ \phi(\lambda),
X^\pm>\vert,
\end{eqnarray}
where $\Delta$ is the Laplacian, in the sense of distributions
(remark that the function $\ln \vert <R^n\circ \phi(\lambda),
X^\pm>\vert$ is subharmonic). To understand the asymptotics of
${1\over N^n} \nu_\nn^\pm$ we have to understand the asymptotics
of the subharmonic functions ${1\over N^n}\ln \vert <R^n\circ
\phi(\lambda), X^\pm>\vert$. This is related to the asymptotics of
the Green function and the Green current of $R$. We do not not
want to enter too much into these details here, and we just sum-up
the main results of \cite{Sabot5}. In particular, we refer to
\cite{Sibony1}, \cite{DFavre}, or to the appendix of
\cite{Sabot5}, for the definitions related to pluricomplex
analysis, rational dynamics.

We set
\begin{eqnarray*}
G_n:\pi^{-1}(\lllG)&\rightarrow &\BR\cup\{-\infty\}
\\
X&\mapsto & \ln\|R^n(X)\|,
\end{eqnarray*}
($G_n$ can be defined on all $\aaa_F$, but we are only interested
in its restriction to $\pi^{-1}(\lllG)$). The functions $G_n$ are
plurisubharmonic (which roughly mean that the restriction to any
holomorphic curve is subharmonic, cf. appendix of \cite{Sabot5}).
The first main result of \cite{Sabot5} is the following
convergence result
$$
\lim_{n\to\infty} {1\over N^n} \ln \vert <R^n\circ \phi(\lambda),
X^\pm>\vert = G\circ \phi(\lambda),
$$ where the convergence is in the sense of $L^1_\loc(\BC)$.  This implies the
following formula for $\mu$.
$$
\mu=\lim_{n\to\infty}{1\over N^n}\nu^\pm_\nn={1\over 2\pi} \Delta
G\circ \phi.
$$
The interest of this formula lies in the fact that $G$ contains a
lot of information about the dynamics of the map $g$ (cf. for
example \cite{Sibony1}).
\begin{remark}
This formula can be seen as a generalization of the classical
Thouless formula combined with an explicit expression for the
Lyapounov exponent in terms of the Green function of the map $R$:
indeed, in the example of the self-similar Sturm-Liouville
operator on $\BR$, the function $G\circ \phi$ coincides with the
value of the Lyapounov exponent $\zeta(\lambda)$ of the propagator
of the underlying ODE. Thus, this formula has two components:
first $\zeta(\lambda)=G\circ \phi(\lambda)$ and $\mu={1\over 2\pi}
\Delta \zeta$. This last equality is exactly the Thouless formula.
\end{remark}
In terms of currents, this result has the following meaning. We
denote by $S_n$ the closed positive $(1,1)$-current with potential
$G_n$, i.e. the current given locally by $(S_n)_{|U}=dd^c G_n\circ
s$ for any local holomorphic section $s$ of $\pi$ on $U\subset
\lllG$ (cf. for example \cite{Sabot5}, section A.3). The current
$S_0$ is the restriction to $\lllG$ of the Fubini-Study form on
$\ppp (\aaa_F)$, and hence is a K\"ahler form on $\lllG$. We
define similarly  $S$, the current with potential $G$.

We define the hypervarieties $C^\pm$ by
\begin{eqnarray*}
C^+&=& \{L\in\lll_F^G,\; \; L\cap (\BC^F\oplus 0)\neq \{0\}\},
\\
C^-&=&\{L\in\lll_F^G,\; \; L\cap (0\oplus (\BC^F)^*)\neq \{0\}\}.
\end{eqnarray*}
We denote by $S_n^\pm$ the closed positive $(1,1)$ current on
$\lllG$ with potential
\begin{eqnarray*}
&&X\mapsto \ln\vert <R^n(X), X^\pm>\vert
\end{eqnarray*}
on $\pi^{-1}(\lllG)$.
It is clear that $S_0^\pm$ are supported by $C^\pm$.
(Indeed, if $s$ is any local holomorphic section of $\pi$ on
$U\subset \lllG$, then $C^\pm\cap U=\{L \in U, <s(L),X^\pm>=0\}$.)
\begin{remark}
If the action of $G$ is trivial, i.e. if $\lll_F^G=\lll_F$, then
it is not difficult to check that $S_0^\pm$ is exactly the current
of integration on the hypersurface $C^\pm$. This may be wrong if
$G$ is not trivial, cf. examples.
\end{remark}
The formula for $\nu_\nn^\pm$, (\ref{formule-nun}),
can be rephrased as follows
\begin{eqnarray*} \nu_\nn^\pm&=&(\pi\circ
\phi)^*(S_n^\pm).
\end{eqnarray*}
N.B: $(\pi\circ \phi)^*(S_n^\pm)$ is the pull-back of the current
$S_n^\pm$ by $\pi\circ\phi$, defined using the local potential,
cf. for example \cite{Sabot5}, A.6.
 \ali
Similarly, the formula for $\mu$ can be translated in terms of current by
$$
\mu=(\pi\circ
\phi)^*S.
$$
\begin{remark}
 This formula says that the density of states
is equal to the section of the current $S$ by the curve
$\pi\circ\phi(\lambda)$.
\end{remark}

 \ali
{\it More general boundary conditions.}  We can generalize the
previous results to more general boundary conditions. Let $B$ be
the complexification of a real Lagrangian subspace of $V$. The
Lagrangian subspace $B$ plays the role of boundary condition:
$B=\BC^F\oplus 0$ for Neumann boundary conditions, and $B=0\oplus
(\BC^F)^*$ for Dirichlet boundary conditions. We denote by $C^B$
the hypervariety $C^B=\{L\in \lllG, \; L\cap B\neq \{0\}\}$. Let
$B^\perp=JB$, be the orthogonal (Lagrangian) subspace of $B$, and
$X_B$ a point in $\aaa_F\setminus \{0\}$ such that
$\pi(X_B)=B^\perp$. We denote by $S_n^B$ the closed positive
$(1,1)$ current on $\lllG$ with potential
$$ \ln\vert < R^n(X),X_B>\vert,$$
on $\pi^{-1}(\lllG)$. Clearly, $S_0^B$ is supported by $C^B$. In
\cite{Sabot5}, appendix D, remark A.7, we have proved the
convergence of $(\pi\circ \phi)^*(S_n^{B_n})$ to the density of
states $\mu=(\pi\circ \phi)^*(S)$, for any sequence of boundary
conditions $B_n$. (Actually, this result was proved only for
boundary conditions of the type $B_n=L_{Q_n}$, for a sequence
$Q_n$ in $\symF(\BR)$, but it is not difficult to adapt the proof
to a general sequence $B_n$.)
 \ali
\begin{remark}
 When $B=L_Q$, for a real symmetric operator $Q$, the measure
$(\pi\circ\phi)^*(S_n^B)$ is equal to the counting measure
$\nu_\nn^Q$ of the spectrum of $(Q_{\rho_\nn}, b_\nn)$ with
boundary condition $Q$, i.e. $\nu_\nn^Q$ is the counting measure
of the eigenvalues of the operator associated with the quadratic
form $<(Q_{\rho_\nn}+Q)\cdot, \cdot>$ and the measure $b_\nn$, as
in section 4.2.
\end{remark}

For any $L$ in $\lll_F^G$ we denote by $\rho_n(L)$ the order of
vanishing of $R^n\circ s$ at $L$, for any local holomorphic
section $s$ of the projection $\pi$ on an open subset $U\subset
\lll_F^G$ containing $L$. We set
$$
\rho_\infty(L)=\lim_{n\to\infty} {1\over N^n} \rho_n(L).
$$
\begin{remark}
Remark that if $G'$ is a subgroup of $G$, then $\lll_F^G\subset
\lll_F^{G'}$ and for any real Lagrangian subspace $L\in\lll_F^G$
the value of $\rho_n(L)$ is the same when we consider the map on
$\lll_F^G$ or $\lll_F^{G'}$, since it is equal to $\dim(\tilde
L_\nn\cap W_\nn^o)$ by proposition \ref{p.4.1}, ii).
\end{remark}

It is clear by proposition \ref{p.4.1}, ii), that
$$
\nu_\nn^\ND(\{\lambda\})=\rho_n(L),
$$
and that
$$
\mu^\ND=\sum_{\lambda\in \BC}
\rho_\infty(\pi\circ\phi(\lambda))\delta_\lambda,
$$
where $\delta_\lambda$ is the Dirac mass at the point $\lambda$.
(N.B.: remark that the sum on the right is finite except for a
denumerable set of reals $\lambda$.)

\begin{remark}
Since $\pi\circ\phi$ satisfies the hypothesis of proposition
\ref{p.4.1}, iii), $\rho_n(\pi\circ\phi(\lambda))$ is also the
order of vanishing of $R^n\circ \phi(\lambda)$ at the point
$\lambda$. Hence, $\nu_\nn^\ND(\{\lambda\})$ is equal to the
Lelong number of $(\pi\circ \phi)^*(S_n)$ at $\lambda$ (and also
of $S_n$ at $\pi\circ \phi(\lambda)$). Yet, it is not clear wether
this equality can be pushed to the limit, i.e. whether
$\rho_\infty(\pi\circ\phi(\lambda))$ is the Lelong number of
$(\pi\circ \phi)^*(S)$ at $\lambda$. This would imply that
$\mu-\mu^\ND$ has no atom. (Relations between the Lelong number of
the Green current and multiplicity of the indeterminacy points of
the map have been established in \cite{Favre1}, but they are not
sufficient to get this result.) Actually, this last equality is
proved in \cite{KigamiL}, using completely different arguments,
based on the renewal theorem.
\end{remark}

\subsection{Asymptotic degree, and the dichotomy theorem}
The relations between the maps $g^n$ and $R^n$ is quite subtle and
the problem of the asymptotic degree and the dichotomy theorem are
essentially related to this. The main idea is that $g^n$ does not
always contain all the useful information about $R^n$: when $R^n$
vanishes on a full projective hypervariety of $\pi^{-1}(\BL_F^G)$,
a term can be locally factorized in $R^n$ and this hypervariety of
annulation does not appear in $g^n$. This is responsible for the
phenomenon of decrease of degrees (which is easier to understand
first in the case of rational maps on the projective space, cf.
\cite{Sibony1}, or appendix B of \cite{Sabot5}). The dichotomy
theorem gives the consequences for the spectrum of the operators
$H_\nn$ of the fact that such a factorization appears or not.

We set
$$
\hat I_\nn=\{ L\in \lll_F^G, \;\;\; \tilde L_\nn\cap W^o_\nn\neq
\{0\}\}.
$$
The analytic set $\pi^{-1}(\hat I_\nn)$ is also the set of zeros
of $R^n$. The set $I_n$ of indeterminacy points of $g^n$
corresponds to the analytic subset of zeros of $R^n$ of
codimension bigger than 1 (cf. \cite{Sabot5}, part 4, for more
details). Let $\{D_{n,1},\ldots ,D_{n,k_n}\}$ be the set of
irreducible components of codimension 1 of $\hat I_\nn$ (which
maybe empty) and $c_{n,j}$ be
%the generic dimension of $\tilde L_\nn\cap W_\nn$ on $D_{n,j}$,
%which is also
the generic order of
vanishing of $R^n$ on $D_{n,j}$. We denote by $D_n$ the divisor
$$
D_n=\sum_{j=1}^{k_n} c_{n,j} D_{n,j},
$$
and by $[D_n]$ the current of integration on the divisor $D_n$,
i.e.  $[D_{n}]=\sum_j c_{n,j} [D_{n,j}]$. We proved in
\cite{Sabot5}, part 4, (but it is almost immediate) that
$$
S_n=(g^n)^*(S_0)+ [D_n].
$$
In \cite{Sabot5}, we proved, for the class of self-similar
structures described there
\begin{thm} \label{t.4.dichotomy}
i) If $D_{n_0}\neq 0$ for a $n_0>0$, then
$$
S=\lim_{n\to \infty} {1\over N^n} [D_n].
$$
The current $S$ is a countable sum of currents of integration on
hypervarieties, and
$$
\mu^\ND=\mu
$$
for all choices of $G$-invariant $\rho$ and $b$ on $F$.

ii) If $D_n=0$ for all $n$, then
$$
S=\lim_{n\to \infty} {1\over N^n} (g^n)^*(S)
$$
is the Green current of $g$. In particular, $S$ is null on the
Fatou set of $g$. Moreover, for a generic choice of $G$-invariant
$\rho$ and $b$, we have
$$
\mu^\ND=0.
$$
\end{thm}
\begin{remark}
This equality $\mu^\ND=\mu$ has strong implications on the nature
of the spectrum of the operator defined on some infinite lattices
constructed from $F_\nn$ (cf. \cite{Sabot6}, and section 5).
\end{remark}
\begin{remark}
The class of self-similar structures considered in \cite{Sabot5}
is a bit more restrictive. But, it is clear from the proof of
proposition 4.2 that this theorem still holds in our setting when
the group of symmetries is trivial, i.e. when $\lll_F^G=\lll_F$.
When the group $G$ is not trivial, the result is based on lemma
4.2 of \cite{Sabot5}, which depends on the particular self-similar
structure we defined there.
\end{remark}

{\it Asymptotic degree.} When $G$ is trivial, then the $(1,1)$
Dolbeault cohomology group of $\lll_F^G=\lll_F$ is of dimension 1
and the degree of $g^n$ is defined as the integer $d_n$ such that
$$
(g^n)^*(\{S_0\})= d_n\{S_0\}.
$$
where $\{S_0\}$ is the cohomology class of the current $S_0$.
Since $\{S_n\}=N^n\{S_0\}$, we have $d_n=N^n$ if $D_n=0$, and
$d_n<N^n$ if $D_n\neq 0$. The asymptotic degree is defined by
$$
d_\infty =\lim_{n\to \infty} (d_n)^{1/n}.
$$
N.B. The sequence $d_n$ is submultiplicative, cf. \cite{Sabot5}.
In \cite{Sabot5}, there is a mistake and $d_\infty$ is wrongly
defined by $\lim {1\over n} \ln d_n$ (which is obviously equal to
$\ln d_\infty$).
 \ali\ali
The case i) of the previous theorem holds when $d_\infty <N$, and
the case ii) holds when $d_\infty =N$.

When $G$ is not trivial, then the $(1,1)$ Dolbeault cohomology
group of $\lll_F^G$ is of dimension $r+1$ ($r+1$ is the number of
irreducible representations of $G$ contained in $\BR^F$). The
degree of the map $g^n$ can be represented by a matrix with
positive integral coefficients. But the asymptotic degree can be
defined anyway, and the same result holds, cf. \cite{Sabot5},
section 4.3.
\begin{remark}
This asymtotic degree is also the asymptotic degree  of some
rational maps, birationally equivalent to $g$, defined on $\BP^k$,
cf. section 4.5 of \cite{Sabot5}. These maps on $\BP^n$ are useful
for practical reasons, since they are sometimes easier to compute.
\end{remark}

\subsection{Spectral analysis of continuous self-similar
Laplace operators}

In general, the finite self-similar structures defined in section
4.1, come from a finitely ramified self-similar set $X$, cf.
section 5.1. Under some conditions, cf. \cite{Kigami1},
\cite{Sabot1}, there exists a (unique) natural self-similar
Dirichlet form $a$ and self-similar measure $m$ on $X$. The
previous results can be generalized to the spectrum of the
infinitesimal generator associated with $(a,m)$. In this case, the
function $\phi$ defined in section 4.6 must be replaced by the
meromorphic function $\phi:\BC\rightarrow \aaa_F$ given by
$$
\phi(\lambda)=\exp\oeta A_{(\lambda)}\eta,
$$
where $A_{(\lambda)}$ is defined as the trace of the Dirichlet
form $a(f,g)+\lambda \int_X f g dm$ on $X$, cf. \cite{Sabot5},
section 3.3.

The crucial but simple fact we want to emphasize here is that
$\pi\circ\phi:\BC\rightarrow \lll_F^G$ is a holomorphic function,
which again satisfies the hypothesis of proposition \ref{p.4.1},
iii).

\section{Post-critically finite self-similar sets and
generalizations}
\subsection{Post critically finite self-similar sets}
We briefly recall the definition of p.c.f self-similar sets of
Kigami (cf. \cite{Kigami1}) to explain where the finite
self-similar structures defined in section 4 appear.

A self-similar set is a compact metric set $X$, together with a
family $(\psi_1, \ldots ,\psi_N)$ of $N$ $\kappa$-Lipchitz
functions from $X$ to $X$, for a $\kappa<1$, and such that
$$
X=\cup_{i=1}^N \psi_i(X).
$$
It is clear, and well-known, that for any sequence
$(i_k)_{k=1}^\infty \in \unN^\BN$, the limit
$$
\lim_{n\to \infty } \Psi_{i_1}\circ \cdots \circ \Psi_{i_n} (x),
$$
converges and that the limit does not depend on the particular
choice of the point $x$. We denote by $\pi((i_k))$ this limit.
Thus we have the following commutative diagram
$$
\begin{CD}
{\unN^\BN} @>\mathrm{\tau_i}>> {\unN^\BN}\\
@VV{\mathrm{\pi}}V @VV{\mathrm{\pi}}V\\
{X}@>\mathrm{\Psi_i}>> {X}
\end{CD}
$$
where $\tau_i$ is the map of $\unN^\BN$ given by
$$\tau_i((i_1,
\ldots ,i_n,\ldots))=(i,i_1,\ldots ,i_n,\ldots ).
$$
We call critical set the set $C\subset \unN^\BN$  given by
$$
C=\cup_{i\neq j} \pi^{-1}(\Psi_i(X)\cap \Psi_j(X)),
$$
and post-critical set the set
$$
P=\cup_{n>0} \sigma^n(C),
$$
where $\sigma $ is the shift map $\sigma((i_1,i_2,\ldots ))=(i_2,
\ldots)$. The set $X$ is called post-critically finite if $P$ is
finite. In this case we set
$$
F=\pi(P),
$$
and it is clear that $F$ satisfies
\begin{eqnarray*}
&&\Psi_i(X)\cap \Psi_j(X)=\Psi_i(F)\cap \Psi_j(F),\;\;\;\forall
i\neq j,
\\
&& F\subset \cup_{i=1}^N \Psi_i(F).
\end{eqnarray*}
This naturally induces a finite self-similar structure if we
define $\rrr$ as the equivalence relation on $\unN\times F$ given
by
$$
(i,x)\rrr(j,y) \;\;\hbox{ iff }\;\; \Psi_i(x)=\Psi_j(y).
$$
The set $\cup_{i=1}^N\Psi_i(F)$ can be clearly identified with
$F_\un =\unN\times F/\rrr$, and contains $F$ as a subset. This
gives a natural identification of a subset of $\pF_\un\subset
F_\un$ with $F$.

We set
$$
F^{(n)}=\cup_{j_1,\ldots ,j_n=1}^N \Psi_{j_1}\circ \cdots \circ
\Psi_{j_n}(F),
$$
which can clearly be identified with $F_\nn$, as defined in
section 4.1. The sequence $F^{(n)}$ is an approximating sequence
of $X$.\ali\ali
 {\it Blow-up of the structure.}\ali
 To simplify the notations,
we suppose here that $X\subset \BR^d$ and that $\Psi_1,\ldots
,\Psi_N$ are defined on  all of $\BR^d$, injective and
$\kappa$-Lipchitz (but its is not necessary, cf. \cite{Sabot5}).

We fix an element $\w$ of $\unN^\BN$, called the blow-up.  For
$n\in \BN$ we set
$$
X_\nn(\w)= \Psi_{\w_1}^{-1}\circ \cdots \circ \Psi_{\w_n}^{-1}(X),
\;\; \partial X_\nn(\w)= \Psi_{\w_1}^{-1}\circ \cdots \circ
\Psi_{\w_n}^{-1}(F),
$$
and
$$
F_\nn(\w)= \Psi_{\w_1}^{-1}\circ \cdots \circ
\Psi_{\w_n}^{-1}(F^{(n)}),\;\; \pF_\nn(\w)= \Psi_{\w_1}^{-1}\circ
\cdots \circ \Psi_{\w_n}^{-1}(F).
$$
Clearly, the sequences $X_\nn(\w)$ and $F_\nn(\w)$ are increasing
and we set
$$
X_\infi (\w)=\cup_{n\in\BN} X_\nn(\w), \;\;\;
F_\infi(\w)=\cup_{n\in \BN} F_\nn(\w),
$$
and $\partial X_\infi=\pF_\infi=\cap_{n}\cup_{m\ge n} \pF_{<m>}$.
Let us remark that the structures at finite level $F_\nn(\w)$ (and
$X_\nn(\w)$)  are ``isomorphic" for different $\w$, but the
unbounded lattice $F_\infi(\w)$ (and the unbounded set
$X_\infi(\w)$) are not in general (cf. \cite{Sabot6}).

The self-similar Schr\"odinger operators defined from
$(Q_{\rho_\nn}, b_\nn)$ can be naturally extended to the unbounded
lattice $F_\infi(\w)$ (cf. \cite{Sabot6}). The measures $\mu$ and
$\mu^\ND$ play an important role in relation with the spectral
properties of these operators. In particular, we have proved in
\cite{Sabot6} that $\supp \mu$ is the spectrum of the operator for
almost all blow-up $\w$ (actually, this is true for any $\w$ such
that $\pF_\infi(\w)=\partial X_\infi(\w)=\emptyset$), and that the
equality $\mu=\mu^\ND$ implies that for almost all $\w$, the
spectrum on $F_\infi(\w)$ is pure point with compactly supported
eigenfunctions. In \cite{Sabot-review1}, we have given some examples
where the role of the blow-up sequence $\w$ is  crucial
with respect to the spectral properties of the operator.

The map $g$ we defined in section 4 plays also an important role
in the construction of a self-similar Dirichlet form on $X$.
Indeed, the subset $\ddd_F^0$, cf. section 1.1, is left invariant
by $g$ and the existence of a non-degenerate self-similar
Dirichlet form on $X$ (cf. \cite{Sabot1} for the appropriate
definitions) is related to the existence of a fixed point of $g$
in the set of irreducible elements of $\ddd_F^0$.

\subsection{Generalization to weighted self-similar operators and
weak connections}
\subsubsection{Weighted self-similar operators}
Let $(\alpha_1,\ldots ,\alpha_N)$ and $(b_1,\ldots ,b_n)$ be two
$N$-tuples of positive reals. We could generalize the previous
setting by considering weighted electrical networks on
${F_\nn}$, obtained by defining $\tilde \rho_\nn$ by
$$
(\tilde \rho_\nn)_{|F_{\nn,i_1,\ldots ,i_n}}= (\alpha_{i_1}\cdots
\alpha_{i_n})^{-1} \rho,
$$
instead of formula (\ref{equa-rho}). Similarly, $\tilde b_\nn$ is
defined on $\tilde F_\nn$ as the sum of the measures $(b_{i_1}
\ldots b_{i_n}) b$ on $F_{\un,i_1,\ldots ,i_n}$, and then
$b_\nn$ is the image of $\tilde b_\nn $ by the projection
$\tilde F_\nn\rightarrow F_\nn$.

Then, we make the following hypothesis.
 \ali

(H) The values $\gamma_i=(\alpha_ib_i)^{-1}$ does not depend on
$i$. We denote by $\gamma_i$ the common value of the $\gamma_i$.

 \ali
Under this hypothesis, the Schr\"odinger operator $H_{\nn,\rho,b}$
associated with $(\rho_\nn,b_\nn)$ is ``locally invariant by
translation" (cf. \cite{Sabot5}).

\subsubsection{Weak connections}
In the definition of section 4.1, the set $F_\nn$ is defined as a
quotient $\tilde F_\nn=\unN^n\times F$. From the electrical point
of view, this means that we add an infinite conductance between
the points of $\tilde F_\nn$ which are connected in $F_\nn$. We
could instead, connect different points in $\tilde F_\nn$ by
positive, but finite, conductance, or mix the two types of
connections. This generalization seems to be useful for
applications, in particular, it seems that Schreier graphs of
certain automatic groups belong to this setting (cf. \cite{Grig1},
\cite{Grig2}).

Formally, this means that we fix a certain electrical network
(eventually dissipative) $\hat \rho_1$ on $\tilde F_\un$, and an
equivalence relation $\rrr$ on $\tilde F_\un$. We define $\tilde
\rho_\un$ as previously, and then $\rho_\un$ is defined as the
electrical network
on ${F_\un}$ obtained by gluing from $\tilde \rho_\un+
\hat \rho_1$.

\begin{remark}
When we introduce weak connection, we loose the homogeneity
property. Indeed, it is no longer true that if we change $\rho$
into $\lambda \rho$, then $\rho_\un$ is changed in $\lambda
\rho_\un$. Similarly, for the renormalization map $g$, the
invariance property (\ref{f.4.1homogeneity}) is no longer valid.
\end{remark}

\subsubsection{Addition and scaling}
To extend the construction of $g$ and $R$ to the case of weighted
self-similar operators and to the case of weak connections, we
have to consider two operations on electrical networks that we
describe now.

 Let $\alpha$ be a non-zero complex number. The linear
operator $Q\mapsto \alpha Q$ on $\sym_F(\BC)$, can be continued to
the compactification $\lll_F$ by the linear operator on $V$ given
by blocks by
$$\tau_\alpha =\left(\begin{array}{cc} \Id & 0\\ 0&\alpha
\Id\end{array}\right).
$$
(Indeed, clearly $\w(\tau_\alpha X, \tau_\alpha Y)=\alpha
\w(X,Y)$, and thus, for $\alpha\neq 0$,  $\tau_\alpha$ acts on
Lagrangian subspaces of $V$.) On the Grassmann algebra,
$\tau_\alpha$ can be lifted by the linear map, that we also denote
$\tau_\alpha$, defined on monomials by
$$
\tau_\alpha(\oeta_{i_1}\eta_{j_1}\cdots \oeta_{i_k}\eta_{j_k})=
\alpha^k \oeta_{i_1}\eta_{j_1}\cdots \oeta_{i_k}\eta_{j_k},
$$
i.e. we have $\tau_\alpha(\pi(X))=\pi(\tau_\alpha(X))$, for any
$X$ in $\pi^{-1}(\lll_F)$.

In the case of weighted self-similar operators, we have to define
$\tilde Q_\un$ as the block diagonal operator on $\tilde F_\un$
equal to $\alpha_i Q$ on the block $\{i\}\times F$. Hence, on the
Lagrangian compactification it means that $\tilde L_\un$ is
defined by
$$
\tilde L_\un =\tau_{\alpha_1}(L)\oplus \cdots \oplus
\tau_{\alpha_N}(L),
$$
with obvious notations. The map $R$ is defined as in the usual
case, except that $\tilde X_\un$ is defined by
$$
\tilde X_\un =(\tau_{\alpha_1}(X))_{\un,1}\cdots
(\tau_{\alpha_N}(X))_{\un,N},
$$
where $(\tau_{\alpha_i}(X))_{\un,i}$ is the copy of
$\tau_{\alpha_i}(X)$ on $\aaa_{\tilde F_{\un,i}}$.

Let us now consider an element $Q_0$ of $\sym_F(\BC)$, and the
linear operator $Q\mapsto Q+Q_0$ on $\sym_F(\BC)$. It is clear
that this operator can be extended to $\lll_F$ by the symplectic
transformation
$$
\tau_{Q_0}=\left( \begin{array}{cc} \Id & 0\\ Q_0&
\Id\end{array}\right).
$$
(Indeed, we have $\tau_{Q_0}(L_Q)=L_{Q+Q_0}$.) On
$\pi^{-1}(\lll_F)\subset \aaa_F$, this operation is lifted by the
multiplication
$$
X\mapsto \exp(\oeta Q_0\eta) X.
$$
In the case of weak connexions, the definition of $g$ and $R$ must
be modified as follows:
 $$ g(L)=\tilde t_{F_\un\rightarrow\pF_\un}\circ
\tilde t_{\tilde F_\un\rightarrow F_\un} (\tau_{Q_{\hat \rho_1}}(\tilde
L_{\un})),
$$
 and
$$
R(X)=R_{F_\un\rightarrow\pF_\un}\circ R_{\tilde F_\un\rightarrow
F_\un} (\exp(\oeta {Q_{\hat \rho_1}}\eta) \tilde X_{\un}).
$$

\section{A class of rational maps on $\lll_K$}
In view of the construction of the map $g$, it is natural to
introduce the following class of rational maps, which contains all
the previous examples and share the same basic properties.
Unfortunately, we know nearly nothing about the dynamics of these
maps.

We consider $V=\BC^K\oplus (\BC^K)^*$, equipped with its canonical
symplectic form $\w$, and $N\in \BN$, $N> 1$. We denote be $\tilde
V_\nn$ the direct sum of $N^n$ copies of $V$:
\begin{eqnarray} \label{f.6.1}
\tilde V_\nn=\oplus_{i_1,\ldots ,i_n=1}^N \tilde V_{\nn,i_1,\ldots
,i_n}.
\end{eqnarray}
We also denote by $\w$ the symplectic structure on $\tilde V_\nn$
induced by $\w$ on $V$.

Let us fix a real coisotropic subspace $W$ of $\tilde V_{<1>}$,
with dimension $(N+1)K$, and denote as usual by $W^o$ its
$\w$-orthogonal subspace. The space $W/W^o$ has dimension $2K$,
and $\w$ induces a symplectic structure on $W/W^o$. We suppose
given an isomorphism (of symplectic structure) between $W/W^o$ and
$V$. Then we define the map $g$ as the composition
$$
g=\tilde t_W\circ (L\mapsto \tilde L_\un),
$$
where, as previously, $\tilde L_\un$ is the Lagrangian subspace of
$\tilde V_\un$ equal to $L\oplus \cdots \oplus L$ for the
decomposition (\ref{f.6.1}), and $\tilde t_W$ is the rational map
defined by the closure of the graph of the symplectic reduction
$t_W$, as defined in section 3. The map $g$ is rational from
$\lll_V$ to $\lll_{W/W^o}\simeq V$.

It is easy to check that the main properties of the map $g$
described in proposition \ref{p.4.1}, remain valid for this class
of maps. In particular, the subset $S_{V,+}$ is invariant by $g$,
and hence is contained in the Fatou set of $g$, since it is
hyperbolic and hyperbolic embedded. The degree of $g$ is smaller
than $N$, and equal to $N$ if and only if the set $\{L, \;\tilde
L_\un\cap W^o\ne \{0\}\}$ has codimension bigger than 1. As
previously, the iterates $g^n$ can also be defined as the
composition of the map $L\mapsto \tilde L_\nn$ and $\tilde
t_{W_\nn}$ for a certain subset $W_\nn\subset \tilde V_\nn$.

It might be interesting to classify this class of maps, up to
isomorphism, in the simplest, but yet non-trivial, case where
$K=2$, and $N=2$. In this case the Lagrangian Grassmannian
$\lll_F$ is 3-dimensional, and isomorphic to the following
subvariety of $\BP^5$
$$
\{[Z,a,d,q,D]\in \BP^5, \;\;\; ad-q^2=DZ\},
$$
cf. \cite{Sabot5}, section 5.2.

\section{Practical implementation and new examples}
It is not always easy to compute the maps $g$ and $R$ since the
dimension of $\aaa_F$ and $\lll_F^G$ might be big, and since
$\lll_F^G$ is not a projective space. The aim of this section is
to describe more precisely than in \cite{Sabot5},  how to proceed,
in practice in a very symmetric case. We suppose here that $\BC^F$
has the following decomposition: $\BC^F=W_0\oplus \cdots \oplus
W_r$, where $W_0, \ldots ,W_r$ are $r+1$ distinct
$\BC$-irreducible representations of $G$, realizable in $\BR$.
(Hence, $\BR^F$ has the same decomposition $\BR^F=W_0\oplus \cdots
\oplus W_r$.) This is the case for example for nested fractals,
\cite{Sabot2}. By Schur's lemma, $\sym_F^G(\BC)\simeq \BC^{r+1}$
and any element $Q$ can be written
$$
Q=u_0p_{|W_0}+\cdots +u_rp_{|W_r},
$$
where $(u_0,\ldots ,u_r)\in \BC^{r+1}$ and $p_{|W_i}$ is the
Hermithian projection on $W_i$. We denote by  $Q^{u_0, \ldots
,u_r}$ the element of $\sym^G_F(\BC)$ of the previous form. The
map $T$ can be represented in coordinates $(u_0,\ldots ,u_r)$;
 it is  rational, and thus can be written under the form
$$
T\left( (u_0, \ldots ,u_r) \right)=\left( {\overline P_0\over
\overline Q_0},\ldots ,{\overline P_r\over \overline Q_r }\right),
$$
where $\overline P_i$, $\overline Q_i$ are polynomials in the
variables $(u_0, \ldots ,u_r)$ (in the usual case of strong
connections, i.e. except in the case of section 5.2.2, the
fraction ${\overline P_i\over \overline Q_i}$ is homogeneous of
degree 1 in the variables $(u_0,\ldots ,u_r)$).
\begin{remark}
The map $T$ is in general not too difficult to compute, even by
hand, since it is just a minimization, and since symmetry
arguments can reduce the number of parameters.
\end{remark}
It is easy to check that, in this case,
the compactification is $\lll_F^G\simeq \BP^1\times \cdots \times \BP^1$. It means
that each coordinates $u_i\in \BC$ is compactified in $\BP^1$.

The compactification $g$ of $T$ on $\BP^1\times \cdots \times
\BP^1$ can be lifted by a polynomial map
$$
\tilde R:\BC^2\times \cdots \times \BC^2\rightarrow \BC^2\times
\cdots \times \BC^2,
$$
of the form
$$
\tilde R\left( (u_0,v_0),\ldots ,(u_r,v_r)\right)= \left( (P_0,
Q_0), \ldots ,(P_r,Q_r)\right),
$$
where $P_i,Q_i$ are polynomials in the variables
$((u_j,v_j))_{j=0,\ldots ,r}$ and homogeneous (with the same
degree) in each of the couple $(u_j,v_j)$, and such that the
following diagram is commutative
$$
\begin{CD}
{\BC^2\times \cdots \times \BC^2} @>\mathrm{\tilde R}>> {\BC^2\times \cdots \times \BC^2}\\
@VV{\mathrm{\pi\times \cdots \times \pi}}V @VV{\mathrm{\pi\times \cdots \times \pi}}V\\
{\BP^1\times \cdots \times \BP^1}@>\mathrm{g}>> {\BP^1\times
\cdots \times \BP^1}
\end{CD}
$$
We denote by $d_{1,i,j}$ the degree of homogeneity of the
polynomials $(P_i,Q_i)$ in the variables $(u_j,v_j)$.

\begin{remark}
 The polynomials $(P_0,Q_0)$ are obtained from the polynomials
$(\overline P_0,\overline Q_0)$ simply by homogeneization (cf.
examples, section 7.2).
\end{remark}
\begin{remark}
These degrees $d_{1,i,j}$ corresponds to the matrix of degrees of
$g$. If we denote by $\nu_i$ the pull-back of the Fubini-Study
form on $\BP^1$ by the projection on the $ith$-factor of
$\BP^1\times \cdots \times \BP^1$, then this means that the
degrees $(d_{1,i,j})$ corresponds to the equation in cohomology
$$
g^*(\{\nu_i\})=\sum_{j=0}^r d_{1,i,j} \{\nu_j\},
$$
where $\{\nu_j\}$ is the cohomology class of $\nu_j$ (the family
$(\{\nu_0\},\ldots ,\{\nu_r\})$ is a basis of the $(1,1)$
cohomology of $\lll_F^G$, cf. \cite{Sabot5}).
\end{remark}

At this point we have just described the map $g$, but this is not
enough to describe the currents $S_n^\pm$ and $S_n$, if there is a
non trivial divisor $D_1$. Let us come back to $\aaa_F$ now. We
have the canonical projections $\pi\times \cdots \times
\pi:\BC^2\times \cdots \times \BC^2\rightarrow \BP^1\times \cdots
\times \BP^1$ and $\pi:\pi^{-1}(\lll_F^G)\rightarrow \lll_F^G$. We
describe a map that lifts the isomorphism $\lll_F^G\simeq
\BP^1\times\ldots\times \BP^1$.  We set $p_i=\dim(W_i)$, and we
denote by $(f_i^1, \ldots ,f_i^{p_i})$ a real orthonormal basis of
$W_i$. Each $f_i^k$ can be written $f_i^k=\sum_{x\in F} c_x e_x$
and we denote by $\ksi_i^k=\sum_{x\in F} c_x \eta_x$ and
$\overline\ksi_i^k=\sum_{x\in F}c_x \oeta_x$ the correponding
vectors in the Grassmann algebra generated by $(\oeta_x, \eta_x)$.
We denote by $\hat s:\BC^2 \times \ldots \times \BC^2 \rightarrow
\aaa_F$ the map given by
$$
\hat s\left( (u_0,v_0),\ldots ,(u_r,v_r)\right)=\prod_{i=0}^r
\prod_{k=0}^{p_i}(v_i+u_i\oksi_i^k\ksi_i^k).
$$
It is clear with these notations that
$$
\hat s \left( (u_0,1),\ldots ,(u_r,1)\right)= \exp\oeta
Q^{u_0,\ldots ,u_r}\eta,
$$
and that $\hat s $ takes its values in $\pi^{-1}(\lll_F^G)$. The
map $\hat s$ is clearly polynomial and homogeneous in each couple
of variables $(u_j,v_j)$ with degree $p_j$, i.e.
$$
\hat s\left( \lambda_0(u_0,v_0),\ldots ,\lambda_r(u_r,v_r)\right)=
\left( \prod_{j=0}^r \lambda_j^{p_j}\right) \hat s \left(
(u_0,v_0),\ldots ,(u_r,v_r)\right).
$$
Since $R$ is polynomial homogeneous of degree $N$, then $R^n\circ
\hat s$ is polynomial homogeneous of degree $N^n p_j$ in
$(u_j,v_j)$.
\begin{remark}
This means, in particular, that the cohomology class of
$S_0$, the current with potential $\ln\|X\|$ on
$\pi^{-1}(\lll_F^G)$, is $\{S_0\}=\sum_{j=0}^r p_j\{\nu_j\},$ and
that $\{S_n\}=N^n\{S_0\}= \sum_{j=0}^r N^n d_j\{\nu_j\}$.
\end{remark}

Since $\tilde R$ and $R$ induce the same map on $\BP^1\times
\cdots \times \BP^1$ then $R\circ \hat s$ is of the form
$$
R\circ \hat s=H \hat s\circ\tilde R,
$$
where $H$ is a polynomial homogeneous in each couple $(u_j,v_j)$,
and we denote by $h_{1,j}$ the homogeneity degree of $H$ in
$(u_j,v_j)$. The divisor $D_1$ is then the divisor associated with
the zeros of $H$. Let us remark that, by homogeneity, these
degrees must satisfy
\begin{eqnarray}\label{f.7.degreef}
N p_i=(\sum_{j=0}^r d_{1,i,j} p_j) + h_{1,i}.
\end{eqnarray}
In practice, the strategy is first to compute $T$, which is in
general quite simple. This gives us the rational map $\tilde R$ by
homogenization. Then, to have all the information about $R$, we
must compute the polynomial $H$, which gives the hypersurfaces of
zeroes of $R$. Either we can guess what are the hypersurfaces of
$\lll_F^G$ where $\tilde L_\un \cap W^o\neq \{0\}$, with
multiplicity, and then check that we found all the factors of $H$
thanks to formula (\ref{f.7.degreef}) (in simple examples it is
quite easy, using symmetry arguments). Or we can use formula
(\ref{f.2.RFpF}), and compute $\det(Q_{|F_\un\setminus
\pF_{\un}})$ in coordinates $(u_0, \ldots ,u_r)$.
\begin{remark}
Formula (\ref{f.7.degreef}) corresponds to the following
equation in cohomology: $\{S_1\}=g^*(\{S_0\}) +\{[D_1]\}.$
\end{remark}
Once $\tilde R$ and $H$ are computed, we have just to iterate
$\tilde R$. At this step, one just has to compute the iterates of
$\tilde R$. The iterates $\tilde R^n$ can be written
$$
\tilde R^n=\left(\tilde H_{n,0}\times (P_{n,0}, Q_{n,0}), \ldots ,
\tilde H_{n,r}\times (P_{n,r},Q_{n,r})\right),
$$
where $H_{n,j}$ are polynomials, homogeneous in the variables
$(u_j,v_j)$, and  where each of the $(P_{n,i}, Q_{n,i})$ are
polynomials with no common factors, and homogeneous in each couple
$(u_j,v_j)$ with the same degree of homogeneity that we denote
$d_{n,i,j}$. This matrix of degrees $(d_{n,i,j}) $ corresponds to
the degrees of $g^n$. We set
$$
\tilde R_n=\left( (P_{n,0}, Q_{n,0}), \ldots ,
(P_{n,r},Q_{n,r})\right).
$$
Then clearly $R^n\circ \hat s$ can be written
\begin{eqnarray*}
R^n\circ \hat s &=& (\prod_{k=0}^{n-1} (H\circ \tilde
R^k)^{N^{n-k}}) \hat s\circ \tilde R^n
\\
&=& (\prod_{k=0}^{n-1} (H\circ \tilde R^k)^{N^{n-k}})
(\prod_{i=0}^r \tilde H_{n,i}^{p_i}) \hat s\circ\tilde R_n.
\end{eqnarray*}
We denote by $H_n$ the polynomial in factor in the last
expression. The divisor $D_n$ is the divisor of $H_n$.

Let us now describe explicitly the spectrum of the operators
$H_\nn$. We denote by $u_0^\rho, \ldots ,u_r^\rho$ the coordinates
of the initial operator $Q_\rho$. The measure $b$ is necessarily a
uniform measure on $F$ (indeed, $G$ acts transitively on the set
$F$, since the trivial representation $W_0$ has multiplicity 1)
hence up to a constant, $I_b=\Id$. We have $\phi(\lambda)=\hat s
((u^\rho_j+\lambda,1)_{j=0,\ldots ,r})$, and the Neumann spectrum
of $H_\nn$ is equal to the zeros of the polynomial
$$ \lambda \mapsto \left( H_n(\prod_{i=0}^r
P_{n,i}^{p_i})\right)((u_j^\rho+\lambda,1)_{j=0,\ldots ,r}),
$$
counted with multiplicities. The Dirichlet spectrum corresponds to
the zeros of the same polynomials where $P_{n,j}$ is replaced by
$Q_{n,j}$. The Neumann-Dirichlet spectrum is obtained by
considering the order of vanishing of
$$
\lambda \mapsto \left( \vert H_n\vert \prod_{i=0}^r
\|(P_{n,i},Q_{n,j})\|^{p_i}\right)((u_j^\rho+\lambda,1)_{j=0,\ldots
,r} ).
$$

\subsection{The Sierpinski gasket}
In this case the connections are described in the following
figure.
 \ali \ali{\input{fig3-1f.pstex_t}}
 \ali
The group $G$ is the group $G\simeq S_3$ of permutations of $F$.
The subspace $W_0=\BC \cdot 1$ of constant functions and its
orthogonal complement $W_1$ are the 2 $\BC$-irreducible
representations of $G$ contained in $\BC^F$, and they are
realizable in $\BR$. In coordinates $(u_0,u_1)$ we have
$$
 T(u_0,u_1)= 3\left( {u_0u_1\over
2u_0+u_1},{u_1(u_0+u_1)\over 5u_1+u_0}\right),
$$
and thus
\begin{eqnarray*}
&&\tilde R\left( (u_0,v_0),(u_1,v_1)\right)
\\
&=&\left((3u_0u_1,2u_0v_1+u_1v_0), (3u_1(u_0v_1+u_1v_0),
5u_1v_0v_1+u_0v_1^2) \right),
\end{eqnarray*}
which means that the map $g$ is represented in homogeneous
coordinates by
\begin{eqnarray*}
&&g\left( [u_0,v_0],[u_1,v_1]\right)
\\
&=&\left([3u_0u_1,2u_0v_1+u_1v_0], [3u_1(u_0v_1+u_1v_0),
5u_1v_0v_1+u_0v_1^2] \right),
\end{eqnarray*}
($[x,y]$ represents the point of $\BP^1$, corresponding to $(x,y)$
in $\BC^2$).
 \ali
 The matrix of degrees is
 $$
d_1= \left(
\begin{array}{cc}
1&1
\\
1&2
\end{array}
\right).
$$
The polynomial $H$ is, up to a constant,
$$
H=v_1.
$$
Indeed, the following vector of $V_{\tilde F_\un}=\BC^{\tilde
F_\un}\oplus (\BC^*)^{\tilde F_\un}$
 \ali
\ali
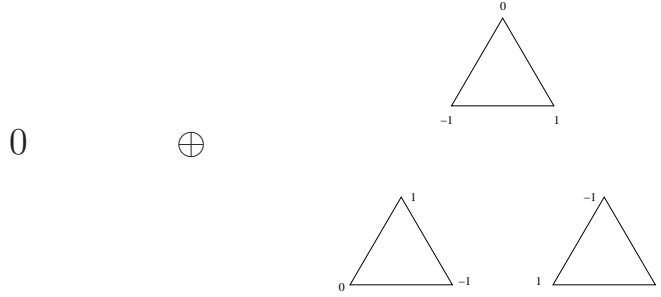
\centerline{\begin{picture}(0,0)%
\includegraphics{gasketfunction.pstex}%
\end{picture}%
\setlength{\unitlength}{1973sp}%
\begingroup\makeatletter\ifx\SetFigFont\undefined%
\gdef\SetFigFont#1#2#3#4#5{%
  \reset@font\fontsize{#1}{#2pt}%
  \fontfamily{#3}\fontseries{#4}\fontshape{#5}%
  \selectfont}%
\fi\endgroup%
\begin{picture}(8251,3630)(2551,-4186)
\put(2551,-2461){\makebox(0,0)[lb]{\smash{\SetFigFont{14}{16.8}{\familydefault}0
}}}
\put(4651,-2461){\makebox(0,0)[lb]{\smash{\SetFigFont{14}{16.8}{\familydefault}$\oplus$
}}}
\end{picture}
}
 \ali
 \ali
 is clearly in $(\tilde
L)_\un\cap W^o$ if $L$ is a Lagrangian subspace in $\lll_F^G$
corresponding to a point of the type $([u_0,v_0],[1,0])$ in
$\BP^1\times \BP^1\simeq \BP^1\times \BP^1$, for any $[u_0,v_0]\in
\BP^1$. Thus, $\dim(\tilde L_\un \cap W^o)$ is generically at
least 1, on the hypersurface $\{v_1=0\}=\BP^1\times [1,0]$.
Equation (\ref{f.7.degreef}) tells us that we found all the
factors of $H$.

\begin{remark}
In \cite{Sabot5}, we described explicitly the current $S$
using a 1-dimensional rational map.
\end{remark}

\subsection{An example coming from group theory}
In  \cite{Grig1}, \cite{Grig2}, Grigorchuk, Bartholdi and Zuk,
considered several examples of fractal groups acting on rooted
trees, and computed their spectrum using a renormalization
equation involving a 2-dimensional rational map. It seems that in
some particular cases, these computation can be performed in our
context, using the generalization of section 5.2.2. We present
here one example: the group $\overline \Gamma$ of \cite{Grig1}. We
don't explain where this example comes from, but just how it can
be described in our context.

The initial cell is 3 points, $F=\{1,2,3\}$. We fix some real
constants $r$ and $v$. Then, we consider the following
self-similar structure, with weak connections, as in section
5.2.2.
 \ali \ali\centerline{\input{grig-new.pstex_t}}
 \ali\ali
This means that in $\tilde F_\un =\tilde F_{\un,1}\sqcup \tilde
F_{\un,2}\sqcup \tilde F_{\un,3}$ we connect the points according
to the previous figure, where the bolded liaisons labelled $r$
represent points connected by a conductance $r$. At each
connecting point we put a dissipative term $v$, as represented on
the figure. This means that if we label the connecting points as
on the following figure
 \ali \ali\centerline{\input{grigexpl.pstex_t}}
 \ali
then $Q_\un=Q_{\un, 1}+Q_{\un, 2}+Q_{\un, 3}+Q_{\hat\rho_1}$,
where $Q_{\hat\rho_1}$ is the matrix associated with the
electrical network $\hat\rho_1$ given by
$$
(\hat\rho)_{x,x'}=(\hat\rho_1)_{x',x''}=\cdots
=(\hat\rho)_{y,y'}=\cdots =r, \;\;
(\hat\rho)_{x}=(\hat\rho_1)_{x'}=\cdots =(\hat\rho)_{y}=\cdots =v.
$$
\begin{remark}
 The example of \cite{Grig1} corresponds to the case
 $r=-t$, $v=2t$, for a real $t$, so that the diagonal terms of $Q_\rho$ are
 cancelled (and actually $t=-1$ in \cite{Grig1}).
\end{remark}

 The circled points represent the
points of $\pF_\un$. It is clear that the structure is invariant
by the group $G$, the group of isometries leaving the triangle $F$
invariant, $G\sim \ddd_3\sim S_3$. As for the Sierpinski gasket,
we have $\BR^F=W_0\oplus W_1$, where $W_0$ is the space of
constant functions and $W_1$ its orthogonal supplement. Any $Q$ in
$\sym_F^G(\BC)$ can be written
$$
Q=u_0p_{|W_0} + u_1p_{|W_1},
$$
where $(u_0,u_1)\in \BC^2$ and $p_{W_0}$ and $p_{W_1}$ are the
orthogonal projections on $W_0$ and $W_1$. A simple computation
gives
\begin{eqnarray}  \label{f.7.Tu}
T((u_0,u_1)) = \left( {3u_0u_1+ z_0 u_0 +2 z_0 u_1\over 2u_0+u_1+3
z_0}, {3u_0u_1+z_1 u_0 +2z_1 u_1\over 2u_0+u_1+3z_1}\right).
\end{eqnarray}
with
$$
z_0=v, \;\;\; z_1=3r+v.
$$
\begin{remark}
In the case of \cite{Grig1}, we have $ z_0=2t$, $z_1=-t$,
which gives
\begin{eqnarray}
T((u_0,u_1))  = \left( {3u_0u_1 +2t u_0 + 4t u_1\over
2u_0+u_1+6t}, {3u_0u_1-t u_0 -2t u_1\over 2u_0+u_1-3t}\right)
\end{eqnarray}
We could compute the map $T$ in different coordinates. For
example, one can represent any $Q$ in $\sym_F^G(\BC)$ in the
following form
$$
Q=\lambda P -\mu \Id,
$$
where $P$ is the matrix null on the diagonal, and equal to 1 on
any off-diagonal term. (NB: this means $P_{i,i}=0$ and
$P_{i,j}=1$, $i\neq j$.) In these coordinates, $T$ has the form
\begin{eqnarray} \nonumber
&&T\left( (\lambda, \mu)\right)
\\ \label{f.7.Tlambda}
&=& \left( {2 \lambda^2 t \over (\lambda-2t-\mu) (\mu-t-\lambda)},
\mu+2\lambda^2{(\lambda-\mu-t)
\over(\lambda-\mu+t)(\lambda-2t-\mu)} \right)
\end{eqnarray}
We remark that this map, when $t=-1$, is exactly the map which
appears in the renormalization equation of lemma 4.14 of
\cite{Grig1}. But this set of variables is not the best suited to
the problem, as we shall see later.
\end{remark}

From equation (\ref{f.7.Tu}), we see that the polynomial map
$\tilde R:\BC^2\times \BC^2\rightarrow \BC^2\times \BC^2$ induced
by $T$ is given by
\begin{eqnarray*}
 \tilde R\left( (u_0,v_0),(u_1,v_1)\right) &=& \left( (3u_0u_1+z_0u_0v_1 +2z_0 u_1v_0,
2u_0v_1+u_1v_0+3z_0 v_0v_1), \right.
\\
 && \left. (3u_0u_1+z_1 u_0v_1 +2z_1 u_1v_0, 2u_0v_1+z_1u_1v_0+3z_1 v_0v_1)
 \right).
\end{eqnarray*}
Thus, the matrix of degrees is
$$
d_1=\left(
\begin{array}{cc}
1 &1 \\1&1 \end{array}\right).
$$
This means in particular that the asymptotic degree $d_\infty$ is
smaller than 2, and that we are in case ii) of theorem
\ref{t.4.dichotomy}.
\begin{remark}
If we compactify in $\BP^2$ then we get a rational map of degree
3. This mean that we cannot see that $d_\infty <3$ at this level
on this compactification. This also means that there will be a
decrease in the degree of the iterates, in this compactification,
cf. section 4.5 of \cite{Sabot5}. Remark also that if we consider
the compactification in $\BP^1\times \BP^1$ for the coordinates
$(\lambda,\mu)$ (which is a complete non-sense), then the degree
of the map, i.e. the largest eigenvalue of the matrix of degrees,
is bigger than 3.
\end{remark}

{\it The divisor $D_1$.}

It is not difficult to check that, up to a multiplicative
constant, the polynomial $H$ is equal to
$$
H=(u_1+z_0)(u_1+z_1)^2.
$$
Indeed, for $u_1=-z_0$ and any $u_0$, it is easy to check that the
following function on $F_\un$ is in $\ker^\ND((Q^{u_0,u_1})_\un)$
 \ali \ali\centerline{\input{grigfunction1.pstex_t}}
 \ali\ali
  Furthermore, for $u_1=-z_1$ and any $u_0$, the following function
 \ali \ali\centerline{\input{grigfunction2.pstex_t}}
 \ali\ali
 and the function obtained by a rotation of $2\pi/3$ are in
$\ker^\ND((Q^{u_0,u_1})_\un)$. Hence, on the hypersurface
$u_1+z_0v_1=0$, $\dim(\tilde L_\un \cap W^o)$ is at least 1, and on
$u_1+z_1v_1=0$ it is at least 2, i.e. we have $h_{1,1}\ge 3$, and
equation (\ref{f.7.degreef}) tells us that there is equality, thus
that we found all the factors of $H$. This implies that the
current $[D_1]$ is equal to
$$
[D_1]= 2[\{u_1+tv_1=0\}]+ [\{u_1-2tv_1=0\}].
$$

\subsection{A semi-symmetric version of the previous example}
We consider now the following example, with weak connection:
 \ali \ali\centerline{\input{grigsemisymm.pstex_t}}
 \ali
As previously, the links labelled $r$ or $r'$, represent
conductances $r$ and $r'$, and the point labelled $v$ and $v'$,
have dissipative terms $v$ and $v'$. We see that now the symmetry
group is no longer $G\simeq S_3$, the group of isometries of the
triangle (since the connecting network is not invariant by
reflections) but the symmetry group $G'\simeq \BZ/3\BZ$ of
rotations of the triangle. But
$\sym_F^G(\BC)=\sym_F^{G'}(\BC)\simeq \BC^2$, since obviously, any
$G'$-invariant $Q$ is of the form
$$
Q^{u_0,u_1}=u_0p_{|W_0}+u_1p_{|W_1}.
$$
A computation gives
\begin{eqnarray*}
T((u_0,u_1))=&(&{3u_0u_1^2+s_0 u_1(2u_0+u_1)+p_0(u_0+2u_1)\over
2u_0u_1+u_1^2+s_0(u_0+2u_1)+3p_0}, \\
&& {3u_0u_1^2+s_1 u_1(2u_0+u_1)+p_1(u_0+2u_1)\over
2u_0u_1+u_1^2+s_1(u_0+2u_1)+3p_1}\;\;),
\end{eqnarray*}
where
\begin{eqnarray*}
s_0=z_0+z'_0, \;\; p_0=z_0z'_0,\;\; s_1=z_1+z'_1,\;\; p_1=z_1z'_1,
\\
z_0=v,\;\; z'_0=v',\;\; z_1=r+v,\;\; z'_1=r'+v'.
\end{eqnarray*}
Indeed, the previous formula is obtained as follows. Consider the
first component of $T$. We need to compute $T(Q^{u_0,u_1})(1)$,
where 1 is the constant function 1 on $F$. The harmonic
continuation of $1$ is, by symmetry, necessarily of the form
 \ali \ali\centerline{\input{grigprol1.pstex_t}}
 \ali
A simple computation gives
\begin{eqnarray}\label{f.ab}
\;\;\;\;\;\; \left\{\begin{array}{l} a={(u_1-u_0)(u_1+z_0')\over d}, \\
b={(u_1-u_0)(u_1+z_0)\over d}, \end{array}\right. \;\;\hbox{where
}\;\; d=2u_0u_1+u_1^2+s_0(u_0+2u_1)+3p_0.
\end{eqnarray}
To compute the second coordinates of $T$, we consider the function
 \ali \ali\centerline{\input{functionW1.pstex_t}}
 \ali
which is in $W_1$ (where $j=e^{2i\pi/3}$). The harmonic
continuation of this function is obviously of the form
 \ali \ali\centerline{\input{grigprolj.pstex_t}}
 \ali
A simple computation gives the same formula for $a$ and $b$ as
formula (\ref{f.ab}), if we replace $z_0,z'_0,s_0,p_0$ by
$z_1,z_1',s_1,p_1$.

 \ali
\begin{remark}
When $r=r'$, and $v=v'$, then we are in the situation of the
previous example, where the group of symmetries is $G\simeq S_3$.
In this case, there are simplifications in both terms of the
formula for $T$. Precisely, the components of $T$ can be written
$$
({(u_1+z_0)( 3u_0u_1+z_0u_0+2z_0 u_1)\over(u_1+z_0)(2u_0+u_1+3z_0)},
{(u_1+z_1)( 3u_0u_1+z_1u_0+2z_1 u_1)\over(u_1+z_1)(2u_0+u_1+3z_1)})
$$
(where we used that $s_0^2=4p_0$ and $s_1^2=4p_1$). We see that we
recover the formulas of 7.2, and that Neumann-Dirichlet
eigenfunctions come from the simplifications in these formulas
(indeed, we can remark that the factors in these equations are
exactly the factors which enter the polynomial $H$: the
multiplicities corresponds to the factor $p_j$, which corresponds
to the dimension of the representation $W_j$, cf. section 7). This
is exactly what is predicted by the general theory.
\end{remark}

We see that the associated homogeneous polynomial on $\BC^2\times \BC^2$
is
\begin{eqnarray*}
&&R((u_0,v_0),(u_1,v_1))
\\
&=&(
(3u_0u_1^2+s_0 u_1(2u_0v_1+u_1v_0)+p_0(u_0v_1^2+2u_1v_0v_1),\\
&&\;\;\; 2u_0u_1v_1+u_1^2v_0+s_0(u_0v_1^2+2u_1v_0v_1)+3p_0v_0v_1^2),\\
&&\;\; (3u_0u_1^2+s_1 u_1(2u_0v_1+u_1v_0)+p_1(u_0v_1^2+2u_1v_0v_1),\\
&&\;\;\;
2u_0u_1v_1+u_1^2v_0+s_1(u_0v_1^2+2u_1v_0v_1)+3p_1v_0v_1^2)).
\end{eqnarray*}
One can easily check that there is no common factor in $R$, hence that the
matrix of degrees is
$$
\left(\begin{array}{cc} 1&1\\ 2&2 \end{array}\right).
$$
Obviously $\left(\begin{array}{c} 1\\2\end{array}\right)$ is an
eigenvector with eigenvalue 3. This means that the polynomial $H$
has degree 0. It means that the current $[D_1]$ is null, hence
that there is no hypersurface of Neumann-Dirichlet eigenvalue at
level 1. We did not check that $[D_n]$ is null for all $n$ (i.e.
that there is no factorization in the iterates $R^n$), but it
seems very probable that it is the case (and it is certainly
possible to verify it, using a formal computation to identify the
contracting curves).

Let us finally mention that the map $f$ on $\BP^1\times \BP^1$, associated with
the homogeneous polynomial $R$ has four indeterminacy points, which are in
$\BC^2$ and given by
$$
(u_0,u_1)=(z_0,z_0),\;(u_0,u_1)=(z'_0,z'_0),\;(u_0,u_1)=(z_1,z_1),\;(u_0,u_1)=(z'_1,z'_1).
$$
We can also remark easily that the diagonal $\{(z,z), z\in \BP^1\}\subset \BP^1\times \BP^1$
is invariant point by point (i.e. $f$ is the identity on the diagonal).

The general theory also predicts that the upper half-plane $\{\im
u_0>0, \im u_1>0\}\subset \BC^2$ is invariant by $f$ (which does
not seem to be easy to check directly on the formulas).


\footnotesize
\begin{thebibliography}{100}
\bibitem{Grig1}
L.  BARTHOLDI and R. I. GRIGORCHUK, {\it On the spectrum of Hecke
type operators related to some fractal groups.} Tr. Mat. Inst.
Steklova 231 (2000), Din. Sist., Avtom. i Beskon. Gruppy, 5--45;
translation in Proc. Steklov Inst. Math. 2000, no. 4 (231), 1--41.
\bibitem{BatesW}
S. BATES and A. WEINSTEIN, {\it Lectures on the Geometry of
Quantization}, Berkeley Mathematics Lecture Notes, 8, 1997.
\bibitem{Berezin} F. A. BEREZIN, The Method of Second
Quantization. Translated from the Russian by Nobumichi Mugibayashi
and Alan Jeffrey. Pure and Applied Physics, Vol. 24 Academic
Press, New York-London 1966 xii+228 pp.
\bibitem{Carlson}
D. CARLSON, {\it What are Schur complements, anyway?} Linear
Algebra Appl. 74 (1986), 257--275.
\bibitem{Colin1}
Y. COLIN DE VERDI\`ERE, {\it R\'eseaux \'electriques planaires I},
Commentarii Math. Helv., 69 (1994), 351-374.
\bibitem{Colin2}
Y. COLIN DE VERDI\`ERE, {\it D\'eterminants et int\'egrales de
Fresnel}, Ann. Inst. Fourier, 49, 3 (1999), 861-881.
\bibitem{Demailly1} J.-P. DEMAILLY,
{\it Monge-Amp\`ere operators, Lelong numbers and intersection
theory}, in Complex Analysis and Geometry, Univ. Ser. Math.,
Plenum Press, pp.115-193, 1993.
\bibitem{DFavre}
J. DILLER and C. FAVRE, {\it Dynamics of bimeromorphic maps of
surfaces},  Amer. J. Math.  123  (2001),  no. 6, 1135--1169.
\bibitem{Favre1}
C. FAVRE, {\it Dynamique des applications rationelles.} Ph. D.
thesis, Universit\'e Paris-Sud-Orsay.
\bibitem{FSibony2}
J. E. FORNAESS, N. SIBONY, {\it Complex dynamics in higher
dimension II.} In Modern Methods in Complex Analysis (Princeton,
NJ, 1992), 135--182, Ann. of Math. Stud., 137, Princeton Univ.
Press, Princeton, NJ, 1995.
\bibitem{Fukushima1}
M. FUKUSHIMA, Y. OSHIMA and M. TAKEDA, Dirichlet Forms and
Symmetric Markov Processes, de Gruyter Stud. Math. 19, Walter de
Gruyter, Berlin, New-york, 1994.
\bibitem{Fukushima2}
M. FUKUSHIMA, {\it Dirichlet forms, diffusion processes and
spectral
 dimensions
for nested fractals,} in : Ideas and Methods in Mathematical
Analysis, Stochastics and Applications, Proc. Conf. in Memory of
Hoegh-Krohn, vol. 1 (S. Albevario et al., eds.), Cambridge Univ.
Press, Cambridge, 1993, pp. 151-161.
\bibitem{GriffithsH}
P. GRIFFITS and J. HARRIS, Principles of Algebraic Geometry. Wiley
Classics Library. John Wiley \& Sons, Inc., New York, 1994.
xiv+813 pp.
\bibitem{Grig2}
R. I. GRIGORCHUK and A. ZUK {\it The lamplighter group as a group
generated by a 2-state automaton, and its
  spectrum.}
Geom. Dedicata 87 (2001), no. 1-3, 209--244.
\bibitem{Hormander}
L. H\"ORMANDER, Notions of Convexity. Progress in Mathematics,
127. Birkhäuser Boston, Inc., Boston, MA, 1994. viii+414 pp.
\bibitem{Kigami1}
J. KIGAMI,  {\it Harmonic calculus on p.c.f. self-similar sets,}
Trans. Am. Math. Soc., 335:721-755, 1993.
\bibitem{KigamiL}
J. KIGAMI and M. L. LAPIDUS, {\it Weyl's problem for the spectral
distribution of Laplacians on p.c.f. self-similar fractals,}
Commun. Math. Phys. 158, no. 1, (1993), 93-125.
\bibitem{Lindstrom}
T. LINDSTR\O M. {\it Brownian motion on nested fractals.} Mem.
Amer. Math. Soc., 420, 1990.
\bibitem{Metz2}
V. METZ, {\it Shorted operators: an application in potential
theory.} Linear Algebra Appl. 264 (1997), 439--455.
\bibitem{Sabot1}
C. SABOT, {\it Existence and uniqueness of diffusions on finitely
ramified self-similar fractals,} in Ann. Scient. Ec. Norm. Sup.,
4\`eme s\'erie, t. 30, 1997, p. 605 \`a 673.
\bibitem{Sabot2}
C. SABOT, {\it Espaces de Dirichlet reli\'es par des points et
application aux diffusions sur les fractals finiment ramifi\'es.}
Potential Analysis, 11 (1999), no. 2, 183-212.
\bibitem{Sabot4}
C. SABOT, {\it Pure point spectrum for the Laplacian on unbounded
nested fractals.}
 J. Funct. Anal. 173 (2000), no. 2, 497--524.
\bibitem{Sabot5}
C. SABOT, Spectral properties of self-similar lattices and
iteration of rational maps,  M\'em. Soc. Math. Fr. (N.S.),  No. 92
(2003), vi+104 pp., arXiv.org/math-ph/0201040.
\bibitem{Sabot6} C. SABOT, {\it Laplace operators on
fractal lattices with random blow-up}, Potential
Analysis, 20, 177-193, 2004. arXiv.org/math-ph/0201041.
\bibitem{Sabot-review1}
C. SABOT,
{\it Spectral analysis of a self-similar Sturm-Liouville
operator}, to appear in Indiana Univ. Math. J., arXiv/math-ph/0401056.
\bibitem{SankaranV}
P. SANKARAN and P.  VANCHINATHAN, {\it Small resolutions of
Schubert varieties in symplectic and orthogonal Grassmannians.}
 Publ.
Res. Inst. Math. Sci. 30 (1994), no. 3, 443--458.
\bibitem{Sibony1}
N. SIBONY, {\it Dynamique des applications rationnelles de $\bold
P\sp k$} (French). In  Dynamique et G\'eom\'etrie Complexes (Lyon,
1997), ix--x, xi--xii, 97--185, Panor. Synthèses, 8, Soc. Math.
France, Paris, 1999.
\bibitem{Siegel}
C. L. SIEGEL, {\it Symplectic Geometry}. Amer. J. Math. 65,
(1943), 1--86.
\bibitem{Wang} J.
SJ\"OSTRAND and W.M. WANG, {\it Exponential decay of averaged
Green functions for random Schr\"odinger operators. A direct
approach.} Ann. Sci. \'Ecole Norm. Sup. (4) 32 (1999), no. 3.
\end{thebibliography}
\end{document}